\documentclass[english]{ws-jai}
\usepackage[T1]{fontenc}
\usepackage[latin9]{inputenc}
\usepackage{array}
\usepackage{float}
\usepackage{multirow}
\usepackage{amsbsy}
\usepackage{graphicx}
\usepackage[flushleft]{threeparttable}
\usepackage{babel}
\floatstyle{ruled}
\newfloat{algorithm}{tbp}{loa}
\providecommand{\algorithmname}{Algorithm}
\floatname{algorithm}{\protect\algorithmname}
\begin{document}

\catchline{}{}{}{}{} 

\markboth{H. Wang}{Combining Multiple Optimised FPGA-based Pulsar Search Modules Using
OpenCL}

\title{Combining Multiple Optimized FPGA-based Pulsar Search Modules Using
OpenCL}
\author{Haomiao Wang, Prabu Thiagaraj, and Oliver Sinnen}


\maketitle

\corres{$^{2}$Corresponding author.}

\thispagestyle{empty}
\begin{abstract}
Field-Programmable Gate Arrays (FPGAs) are widely used in the central
signal processing design of the Square Kilometre Array (SKA) as acceleration
hardware. The frequency domain acceleration search (FDAS) module is
an important part of the SKA1-MID pulsar search engine. To develop
for a yet to be finalised hardware, for cross-discipline interoperability
and to achieve fast prototyping, OpenCL as a high-level FPGA synthesis
approach is employed to create the sub-modules of FDAS. The FT convolution
and the harmonic-summing plus some other minor sub-modules are elements
in the FDAS module that have been well-optimised separately before.
In this paper, we explore the design space of combining well-optimised
designs, dealing with the ensuing need to trade-off and compromise.
Pipeline computing is employed to handle multiple input arrays at
high speed. The hardware target is to employ multiple high-end FPGAs
to process the combined FDAS module. The results show interesting
consequences, where the best individual solutions are not necessarily
the best solutions for the speed of a pipeline where FPGA resources
and memory bandwidth need to be shared. By proposing multiple buffering
techniques to the pipeline, the combined FDAS module can achieve up
to 2x speedup over implementations without pipeline computing. We
perform an extensive experimental evaluation on multiple FPGA boards
(Arria 10) hosted in a workstation and compare to a technology comparable
mid-range GPU. 
\end{abstract}


\section{Introduction}

For a large scale global project such as the Square Kilometre Array~(SKA)~\footnote{www.skatelescope.org},
hundreds of research institutes and companies from over ten member
countries are enrolled~\cite{dewdney2009square}. Each research group
is assigned a small task such as one or several modules of the overall
pipeline. After each module is investigated and optimised, it needs
to be integrated with modules from other groups to form the whole
pipeline. For software designs, different institutes can use the same
operating system such as Linux and development environment. A large
number of programming languages can be applied, and the software developers
only need to make sure the external application programming interface
(API) can be used by other groups. 

Field-programmable gate arrays (FPGAs) and Graphics processing units
(GPUs) are two main types of accelerators in radio astronomy projects.
For GPU development, CUDA and OpenCL can be employed in the development
and the details vary based on the GPU brand. In terms of FPGA development,
the traditional synthesis flow needs hardware description languages~(HDLs)
such as Verilog HDL and VHDL, which is hard to understand let alone
modify for SKA collaborators (e.g., software engineers and physicists)
without expert knowledge in hardware design. Besides the traditional
flow, a large number of high-level synthesis tools support a variety
of high-level languages (compared to HDLs) such as OpenCL~\cite{czajkowski2012opencl},
C/C++, and Java~\cite{maxeler2001wp}.
In the SKA project, a framework that executes across heterogeneous
platforms such as OpenCL is an excellent option for prototyping designs
using acceleration hardware. By applying OpenCL, the same kernel codes
can be executed on both FPGAs and GPUs without substantial modification,
providing the same functionality of the design. While this is very
useful, the performance of a single OpenCL design might vary strongly
across platforms, due to the difference between the structures of
FPGAs and GPUs, and require some 'performance porting' between different
types of devices. The use of OpenCL makes the code more accessible
to non-hardware-designers, provides functional portability and easy
generational upgrades within a device type.

In this research, we investigate the Fourier domain acceleration search~(FDAS)
module~\cite{ransom2001fast} of the pulsar search engine~(PSS)
within the SKA1-MID central signal processor~(CSP). The main function
of the FDAS module is to remove the smeared pulsar signals by using the
correlation technique~\cite{ransom2002fourier,jouteux2002searching}.
It consists of two main parts: FT convolution sub-module and harmonic-summing
sub-module. The FT convolution module is a compute-intensive application
that contains 85 FIR filters, with up to 400 coefficients (or taps).
The harmonic-summing module is a data-intensive application, and the
main problem is the large number of irregular memory accesses during
processing. These two modules have been individually investigated
and well-optimised on high-end FPGAs using OpenCL in previous research\@.
The optimised designs can gain better performance and consume less
energy on FPGAs than that of GPU designs while meeting the requirements.
However, the optimised performance might not be achieved when combining
with other modules. More interestingly, optimisation choices might
be different when sub-modules are part of a larger pipeline. We investigate
in this paper the combination of well-optimised designs, explore the
design space and optimise the combination of designs. The main contributions
of this research are as follows:
\begin{itemize}
\item Design Space: explore the design space of combining investigated implementations;
three types of data transformation methods are investigated to combine
proposed FT convolution and harmonic summing implementations;
\item Pipeline structure: adopting multiple buffering (double and triple
buffering in this research) to improve the performance of investigated
combinations;
\item Multiple Devices: multiple acceleration devices are employed in processing
the combined implementations. Different methods of partitioning the
workload across devices are investigated.
\end{itemize}
The rest of the paper is organized as follows.  Section~\ref{sec:FDAS-Module}
provides the details of straight-forward and optimised designs of
the FT convolution module and the harmonic-summing module and states
the design goals of the FDAS module. In Section~\ref{sec:OpenCL-based-Architecture},
the design space of combining optimised modules is explored, and
the pipeline structure is investigated on multiple devices. Section~\ref{sec:Evaluation}
presents the experimental evaluation results and their analysis. Finally,
the conclusions are given in Section~\ref{sec:Conclusions}.

\section{\label{sec:FDAS-Module}Frequency Domain Acceleration Search}

\begin{figure*}[t]
\begin{centering}
\includegraphics[bb=10bp 0bp 510bp 309bp,clip,scale=1]{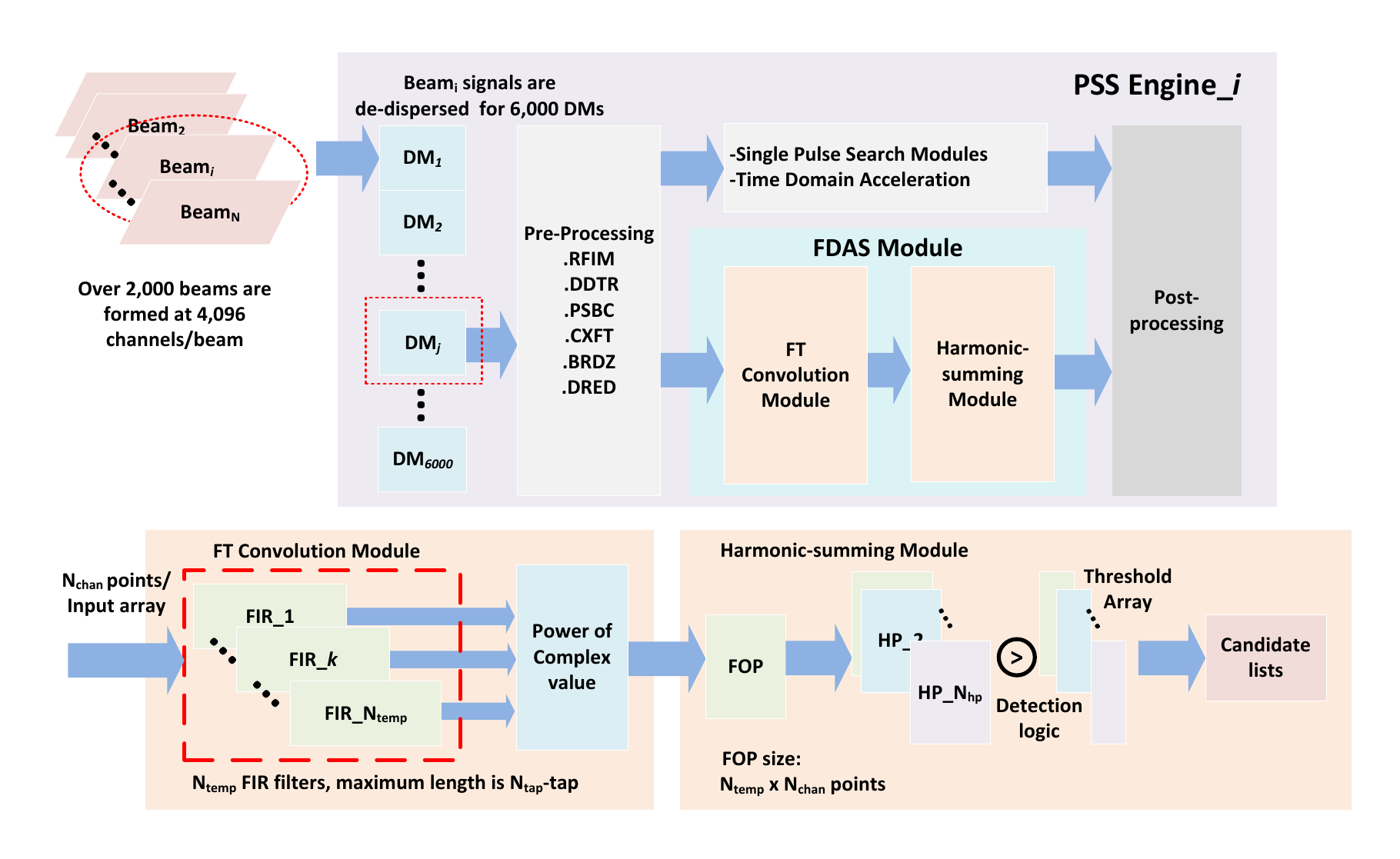}
\par\end{centering}
\caption{\label{fig:The-Processing-Flow}Processing flow of Pulsar Search Engine
(PSS) of SKA1 CSP system and details of FDAS module}
\end{figure*}

The FDAS module illustrated in Figure~\ref{fig:The-Processing-Flow}
is a part of the SKA1-MID CSP element, and the required parameters
are listed in Table~\ref{tab:Harmonic-summing-Module-Paramete}.
From the antennas, over 2,000 beams are formed at 4,096 frequency
channels per beam. The signals of each beam are processed independently,
and each beam needs a dedicated pulsar search engine. Because the
dispersion measure, to compensate for signal changes due to travel
through interstellar space, is unknown, over 6,000 trial values are
tested, and several pulsar search approaches are employed such as
time domain acceleration search and frequency domain acceleration
search. The FDAS module consists of two main parts: 1) the FT convolution
module and 2) the harmonic-summing module. Both these modules 
have been investigated and optimised for FPGAs before (the FT convolution
module in~ \cite{wang2016fpga,wang2018trets} and the Harmonic-summing
module in~\cite{wang2018vlsi}), and we very briefly review the
details in this section. 

\begin{wstable}[h]
\caption{\label{tab:Harmonic-summing-Module-Paramete}FDAS Module Parameters}
\centering{}{\footnotesize{}}%
\begin{tabular}{@{}ccc@{}} \toprule
{\normalsize{}Parameter} & {\normalsize{}Description} & {\normalsize{}Value}\\
\colrule 
{\hphantom{}$N_{beams}$} & {\hphantom{}Number of beams} & {\hphantom{}1000\textasciitilde{}2000}\tabularnewline
{\hphantom{}$N_{DM-trail}$} & {\hphantom{}Number of de-dispersion measure (DM) trails} & {\hphantom{}$6000$}\tabularnewline
{\hphantom{}$T_{obs}$} & {\hphantom{}time period of each observation} & {\hphantom{}$540s$}\tabularnewline
{\hphantom{}$N_{temp}$} & {\hphantom{}Number of templates (row of the FOP )} & {\hphantom{}$85$}\tabularnewline
{\hphantom{}$N_{chan}$} & {\hphantom{}Number of channels (column of the FOP)} & {\hphantom{}$2^{21}$}\tabularnewline
{\hphantom{}$N_{tap}$} & {\hphantom{}Number of FIR filter taps for each template} & {\hphantom{}$421$}\tabularnewline
{\hphantom{}$N_{hp}$} & {\hphantom{}Total number of harmonic planes} & {\hphantom{}$8$}\tabularnewline
{\hphantom{}$N_{cand}$} & {\hphantom{}Number of candidates per harmonic plane} & {\hphantom{}$200$}\tabularnewline
\botrule
\end{tabular}
\end{wstable}

\begin{figure}
\begin{centering}
\includegraphics[bb=20bp 20bp 560bp 350bp,clip,scale=0.7]{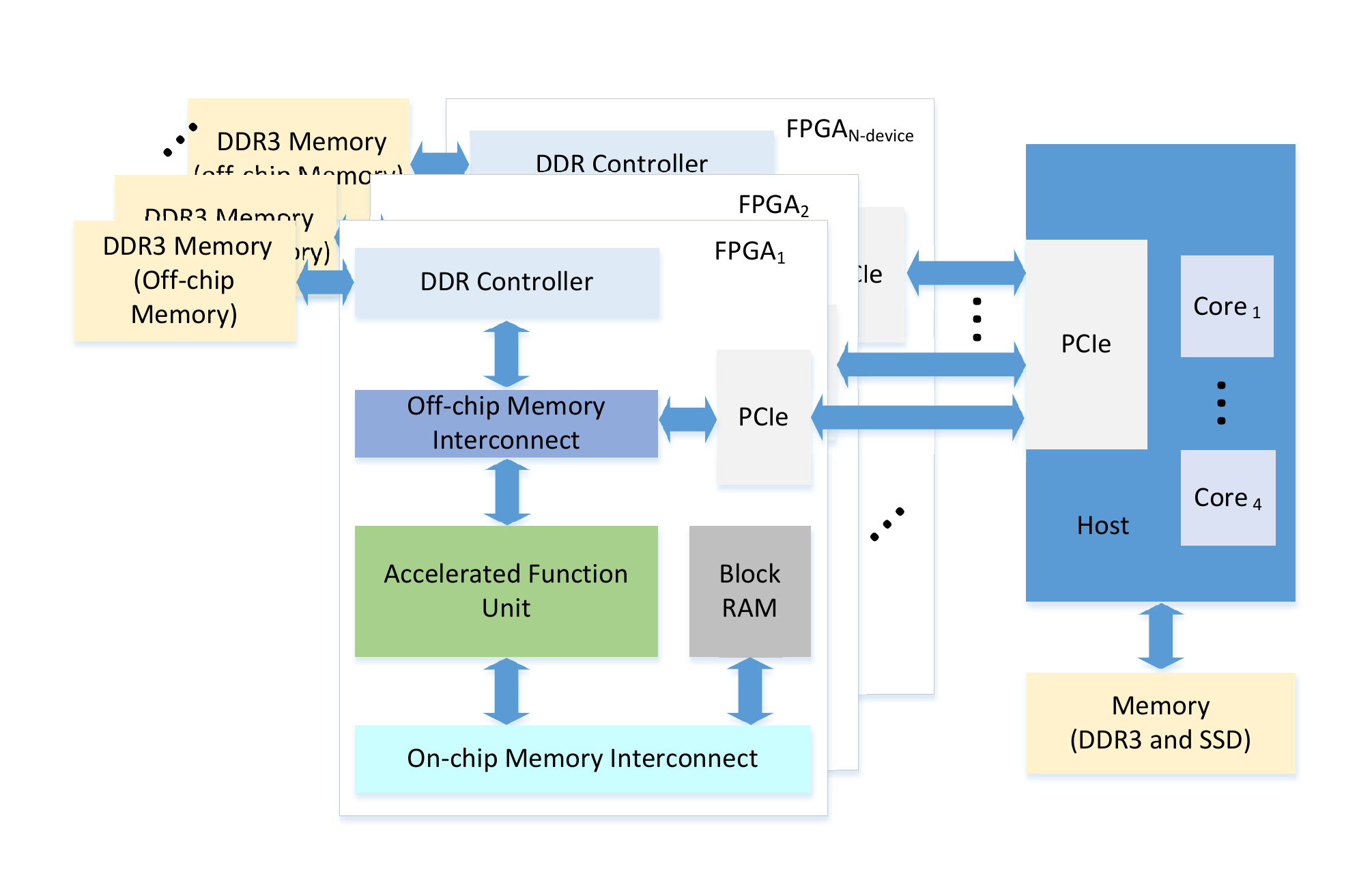}
\par\end{centering}
\caption{\label{fig:FPGA-devices-as}Structure of the high-end FPGA devices
as acceleration hardware in a host system}
\end{figure}

In previous research, different types of acceleration devices were
employed to evaluate the performance of the straight-forward and optimised
approaches. Two types of Intel high-end FPGAs (Stratix V, referred
to as $\mathbf{S5}$, and Arria 10, referred to as $\mathbf{A10}$)
are compared with one mid-range AMD R7 GPU, referred to as $\mathbf{R7}$.
The platform specifications are given in Table~\ref{tab:Details-of-FPGA}.
The FPGA and GPU cards are connected to the host through the PCIe
bus, and the structure of FPGA-based platform is depicted in Figure~\ref{fig:FPGA-devices-as}.
For the FPGA acceleration cards, each one is connected through 8x
lane PCIe bus ($S5$ use PCIe Gen2.0 and $A10$ use PCIe Gen3.0).
Regarding the $R7$ GPU board, it uses a 16x lane PCIe bus of Gen3.0.
Gen3.0.

\begin{wstable}[h]
\caption{\label{tab:Details-of-FPGA}Details of CPU, GPU and FPGA Platforms}
\centering{}{\scriptsize{}}%
\begin{tabular}{@{}cccccc@{}}
\toprule 
\multirow{2}{*}{{\small{}Device}} & {\small{}Terasic DE5-Net } & {\small{}Nallatech 385A } & {\small{}Sapphire Nitro R7 370 } & {\small{}Intel CPU Host }\tabularnewline
 & {\small{}(}\textbf{\small{}S5}{\small{})} & {\small{}(}\textbf{\small{}A10}{\small{})} & {\small{}(}\textbf{\small{}R7}{\small{})} & {\small{}(}\textbf{\small{}I7}{\small{})}\tabularnewline
\colrule  
\multirow{2}{*}{{\small{}Hardware}} & {\small{}Intel Stratix V } & {\small{}Intel Arria 10 } & {\small{}AMD Radeon } & {\small{}Intel Core }\tabularnewline
 & {\small{}5SGXA7} & {\small{}GX1150} & {\small{}R7 370} & {\small{}i7-6700K}\tabularnewline
{\small{}Technology} & {\small{}$28nm$} & {\small{}$20nm$} & {\small{}$28nm$} & {\small{}$14nm$}\tabularnewline
\multirow{2}{*}{{\small{}Compute resource}} & {\small{}622,000 LEs} & {\small{}1,506,000 LEs} & {\small{}1,024 Stream Processors} & {\small{}8 Processors}\tabularnewline
 & {\small{}256 DSP blocks} & {\small{}1,518 DSP blocks} & {\small{}(16 Compute Units)} & {\small{}(4 Cores)}\tabularnewline
{\small{}On-chip memory size} & {\small{}50$Mb$} & {\small{}53$Mb$} & {\small{}\textemdash{}} & {\small{}64$Mb$}\tabularnewline
{\small{}Off-chip memory size} & {\small{}2 x $2GB$ DDR3} & {\small{}2 x $4GB$ DDR3} & {\small{}$4GB$ GDDR5} & {\small{}$64GB$ DDR4}\tabularnewline
{\small{}Memory interface width} & {\small{}2 x 64-bit} & {\small{}2 x 72-bit} & {\small{}256-bit} & {\small{}\textemdash{}}\tabularnewline
{\small{}Max clock frequency} & {\small{}600$MHz$} & {\small{}1.5$GHz$} & {\small{}985$MHz$} & {\small{}4.2$GHz$}\tabularnewline
{\small{}Max power consumption} & {\small{}\textemdash{}} & {\small{}75W} & {\small{}150W} & {\small{}\textemdash{}}\tabularnewline
\botrule 
\end{tabular}
\end{wstable}

Apart from these PCIe card-based platforms, the Intel Xeon Scalable
processor with an in-package Arria FPGA from the Hardware Accelerator
Research Program (HARP) is employed in this research. The platform,
referred to as $HARP$, has a 14 core Xeon processor at $2.4GHz$
and an Intel Arria 10 GX1150 FPGA, which is the same as the one on
the $A10$ card.

\subsection{\label{subsec:FT-Convolution-Module}FT Convolution Module}

The core computation part of the FT convolution module is to process
$N_{chan}$ points with $N_{temp}$ FIR filters. The basic FIR filter
implementation is investigated in both time domain (TDFIR) and frequency
domain (FDFIR).

\subsubsection{Frequency domain \textendash{} FDFIR}

\paragraph{Na\text{\"i}ve TDFIR}

The TDFIR filter is a straight-forward implementation of equation~\ref{eq:tdfir}

\begin{equation}
y[i]=\sum_{j=0}^{N_{tap}-1}x[i-j]h[j],\,for\,i=0,1,\,...N_{chan}-1,\label{eq:tdfir}
\end{equation}
where $x[\cdot]$, $h[\cdot]$, and $y[\cdot]$ are complex single
precision input signals, coefficients, and output results, respectively.

\paragraph{Overlap-add Algorithm based TDFIR}

The amount of logic resources and DSP blocks in a specific FPGA are
fixed. If the FIR filter size $N_{tap}$ is too large, an FPGA does
not have enough logic resources and DSP blocks to parallelise $N_{tap}$
complex multiplications and then fails to achieve a pipeline structure.
To make an $N_{tap}$-tap FIR filter fit into the targeted FPGA and
maintain high-performance, we apply the overlap-add algorithm (\textbf{OLA})
to split the coefficient array into a group of sub-arrays~\cite{pavel2013algorithms}.  

\subsubsection{Frequency domain \textendash{} FDFIR}

\paragraph{Na\text{\"i}ve FDFIR}

Based on the convolution theorem, Equation (\ref{eq:fdfir-1}), the
output of an FIR filter can be obtained by the following three steps~\cite{smith1997scientist}:
1) Fourier transform of the input array and coefficient array, 2)
element-wise multiplication of these two arrays, and 3) inverse Fourier
transform of the output array. 
\begin{equation}
x\ast h=\mathcal{F}^{-1}\{\mathcal{F}\{x\}\cdot\mathcal{F}\{h\}\},\label{eq:fdfir-1}
\end{equation}
where $\mathcal{F}\{\cdot\}$ and $\mathcal{F}^{-1}\{\cdot\}$ are
Fourier transform and inverse Fourier transform.

\paragraph{Overlap-save Algorithm based FDFIR}

For large input size Fourier transforms, such as the targeted two
million points ($2^{21}$) FFT, the on-chip memory of an FPGA is unable
to store all points, which makes it impossible to perform the complete
process as described in Equation (\ref{eq:fdfir-1}) in one go. Hence,
we apply the overlap-save algorithm (OLS) to split the input signals
into chunks~\cite{pavel2013algorithms}. Each chunk overlaps with
its two neighbour chunks, and the extent of the overlap is $N_{tap}-1$.
For the first input chunk, $N_{tap}-1$ zero points have to be padded
at the beginning. After convolving in frequency-domain, the overlap,
which is the first $N_{tap}-1$ points of each chunk, are discarded.

\subsubsection{Optimised Performance}

\begin{figure}
\begin{centering}
\includegraphics[bb=0bp 10bp 504bp 340bp,clip,scale=0.8]{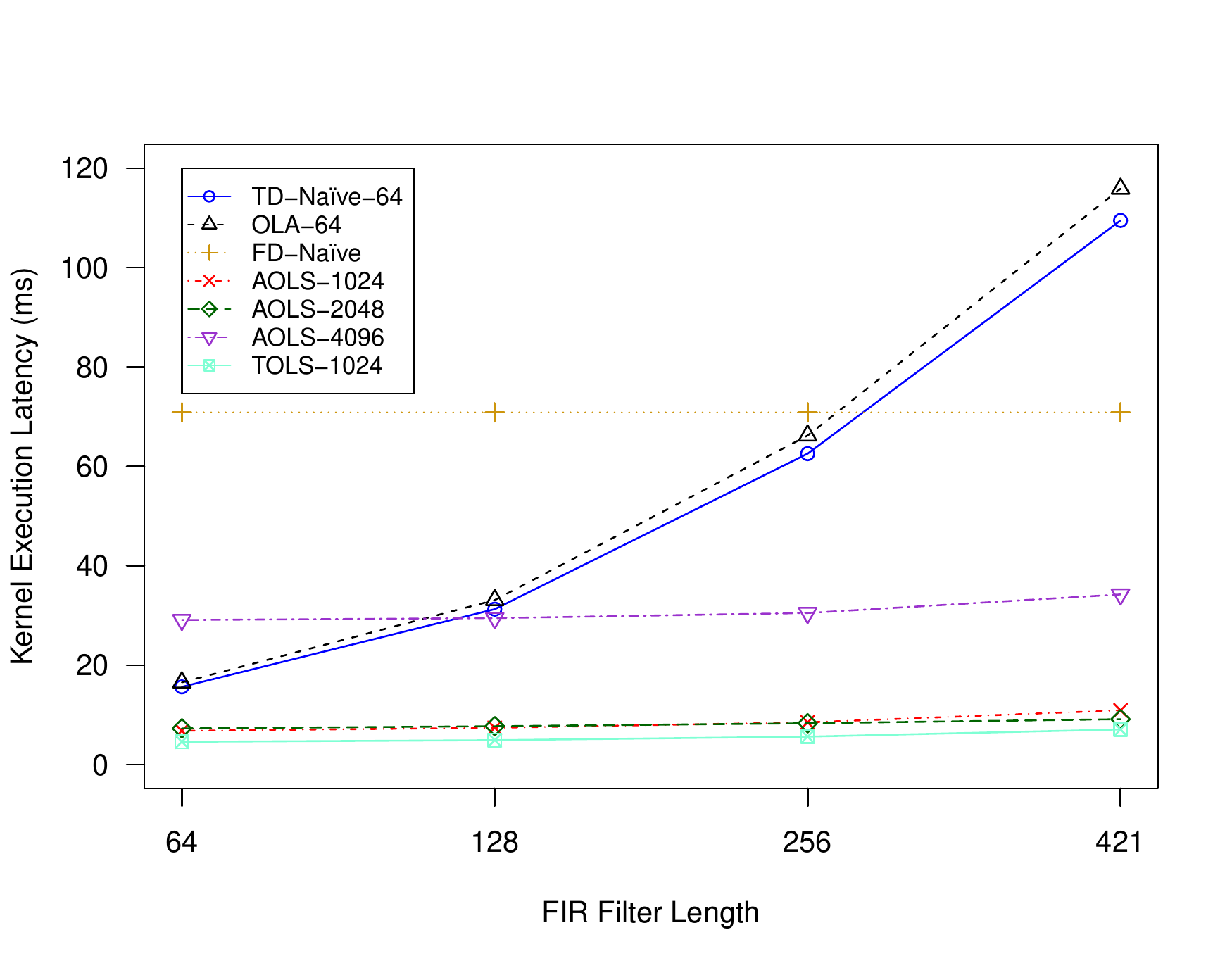} 
\par\end{centering}
\caption{\label{fig:Exe-late-FT}Execution latency of a single FIR filter using
TDFIR and FDFIR based kernels on one $S5$ }
\end{figure}

The straight-forward and optimised implementations of a single FIR
filter are evaluated, and the execution latencies of these kernels are
given in Figure~\ref{fig:Exe-late-FT}. For TDFIR kernels, the value
64 represents a completely parallelised 64-tap FIR filter. The $S5$
FPGA has 256 DSP blocks and 64 complex SPF multiplications are the
largest scale it can parallelise. \textbf{AOLS} is the area-efficient
OLS-FDFIR that contains only one FFT engine (radix-4 feed-forward
FFT~\cite{garrido2013pipelined}). The AOLS kernels have to be launched
twice to process one input array. The \textbf{TOLS} is the time-efficient
OLS-FDFIR that contains two FFT engines. The number after AOLS and
TOLS in the legend of Figure~\ref{fig:Exe-late-FT} indicates the
size of the point chunks. 

The experiments in \cite{wang2016fpga} demonstrated that TOLS-1024
is the fastest among these kernels in implementing \emph{one} FIR
filter. For $N_{temp}$ FIR filters, kernel TOLS-1024 has to Fourier
transform the same input array $N_{temp}$ times. The AOLS kernels
then become efficient since it can Fourier transform the input array
once and then launch $N_{temp}$ times to implement $N_{temp}$ FIR
filters.

\begin{figure}
\begin{centering}
\includegraphics[bb=0bp 10bp 504bp 340bp,clip,scale=0.8]{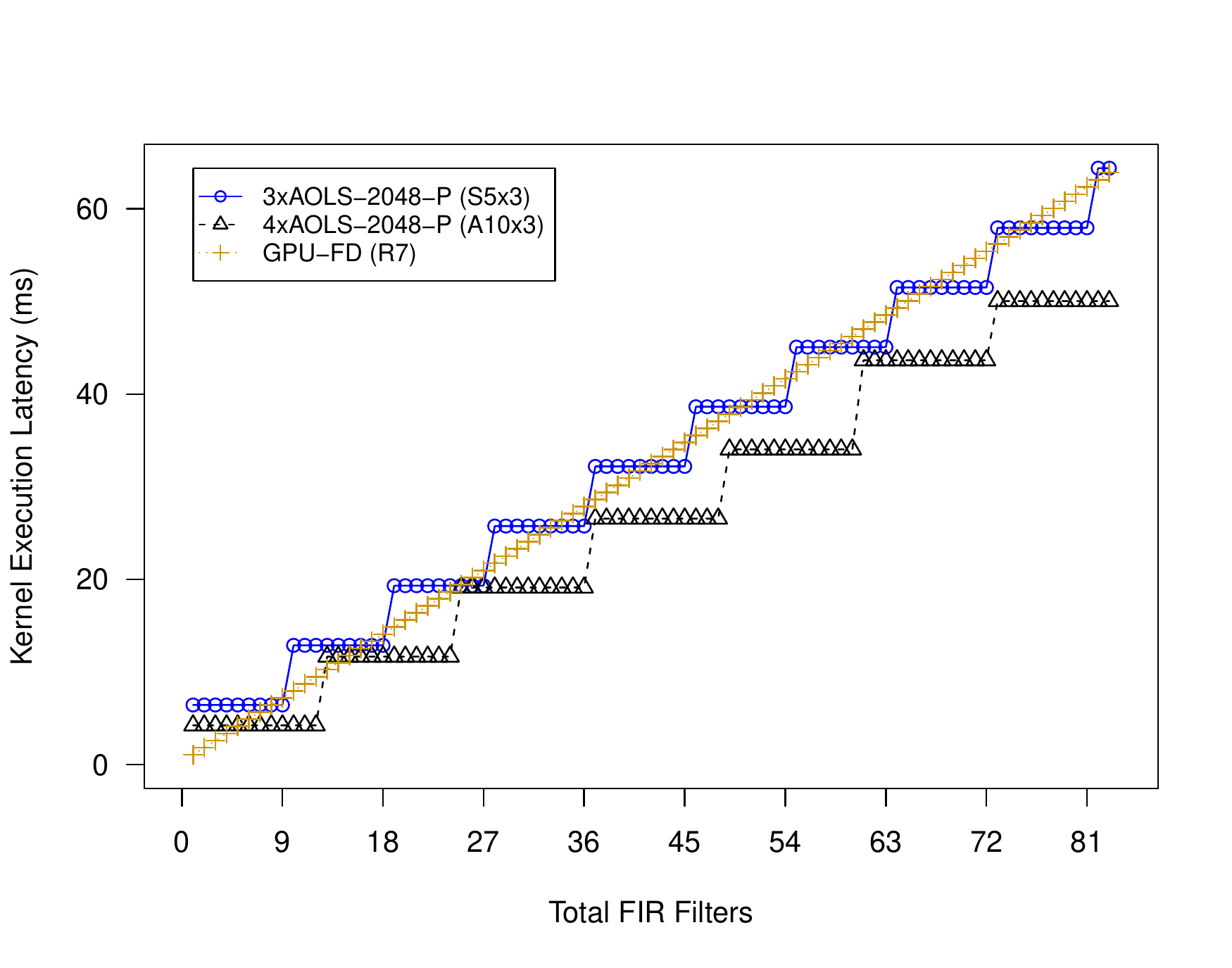}
\par\end{centering}
\caption{\label{fig:Exec-fastest-fir}Execution latency of AOLS-2048 kernel
using three $S5$ and $A10$ FPGAs, and a single $R7$ GPU}
\end{figure}

Using the fastest implementation of the optimised designs (AOLS-2048),
the pipeline is slightly extended to include the power calculation
of the complex filter outputs and then evaluated on two types of FPGA
devices and one GPU device. The results over varying numbers of FIR
filters are given in Figure~\ref{fig:Exec-fastest-fir}. For FPGAs,
the results are based on employing three cards, and the same AOLS-2048-P
kernel can be replicated 3x times on each $S5$ and 4x times on each$A10$.
It can be seen that three $A10$ cards can execute the FT convolution
module in about $50ms$, and it is 1.3x times faster than the single
$R7$ GPU, which uses significantly more power than three FPGA boards. 

\subsection{\label{subsec:Harmonic-summing-Module}Harmonic-summing Module}

The FT convolution output is the Filter-Output-Plane (FOP) that is
sent to the harmonic-summing module for candidate detection. In the
harmonic-summing module, which is described in Algorithm~\ref{alg:General-Harmonic-summing-Algorit},
the FOP is stretched by a group of integers to generate $N_{hp}$
stretch planes ($SP$s). The FOP and stretch planes are accumulated
to calculate $N_{hp}$ harmonic planes ($HP$s) and then threshold-detection
logic is applied to collect $N_{cand}$ candidates from each harmonic
plane. All operations in the harmonic-summing module are inexpensive
operations such as floating-point additions and comparisons with a
constant. The FOP takes up to $710MBytes$ under current requirements,
which is tens of times larger than the on-chip memory size of a high-end
FPGA, so it has to be stored in the off-chip memory (i.e., DDR RAM
on FPGA board) in the FT convolution module. The main issue for this
module is the large number of irregular off-chip memory accesses,
and we optimised this issue using two approaches: 1) reducing the
number of accesses and 2) increasing the used off-chip memory bandwidth~\cite{weinhardt1999memory}. 

\begin{algorithm}
\caption{\label{alg:General-Harmonic-summing-Algorit} Harmonic-summing Algorithm\textsc{ }}

{\small{}$SP_{1}\leftarrow$(filter-output-plane (FOP))}{\small \par}

{\small{}$CL\leftarrow0$ \{initialize the detection output\}}{\small \par}

\textbf{\small{}for $k=1$ }{\small{}to $N_{hp}$ }\textbf{\small{}do}{\small \par}

{\small{}~~~~}\textbf{\small{}for $i=-(N_{temp}-1)/2$ }{\small{}to
$(N_{temp}-1)/2$ }\textbf{\small{}do}{\small \par}

{\small{}~~~~}\textbf{\small{}~~~~for $j=0$ }{\small{}to
$N_{chan}-1$ }\textbf{\small{}do}{\small \par}

{\small{}~~~~~~~~~~~~$SP_{k}(i,\,j)\leftarrow$stretch$(SP_{0},\,k,\,i,\,j)$
\{generate the value in stretched plane\}}{\small \par}

{\small{}~~~~~~~~~~~~$HP_{k}(i,\,j)\leftarrow HP_{k-1}(i,\,j)+SP_{k}(i,\,j)$
\{based on the stretched plane, generate the value in harmonic plane\}}{\small \par}

{\small{}~~~~~~~~~~~~$CL\leftarrow$detection$[HP_{k}(i,\,j),\,TA(k,\,i)]$
\{threshold-detection logic to identify valid peak signals\}}{\small \par}

{\small{}~~~~~~~~}\textbf{\small{}end for}{\small \par}

{\small{}~~~~}\textbf{\small{}end for}{\small \par}

\textbf{\small{}end for}{\small \par}

{\small{}Candidate List $\leftarrow CL$}{\small \par}
\end{algorithm}

Two types of methods for the processing in the harmonic-summing module
were investigated: \textsc{SingleHP}, where a single harmonic plane
is processed at a time, and \textsc{MultipleHP,} where multiple harmonic
planes are processed simultaneously. The optimised methods
are listed below and the parameters are described in Table~\ref{tab:HM-parameters}.
\begin{itemlist}
\item \textsc{SingleHP}
\begin{itemlist}
\item \textsc{SingleHP-$(S/M,\,V/R,\,N_{paral})$}
\end{itemlist}
\item \textsc{MultipleHP}
\begin{itemlist}
\item Na\text{\"i}ve \textsc{MultipleHP}
\item \textsc{MultipleHP-H-$(N_{MultipleHP-H-preld})$}
\item \textsc{MultipleHP-N-$(N_{MultipleHP-N-col})$}
\item \textsc{MultipleHP-R-$(N_{MultipleHP-R-col},N_{points/wi})$}
\end{itemlist}
\end{itemlist}
\begin{wstable}[h]
\caption{\label{tab:HM-parameters}Parameters of harmonic-summing methods}
\centering{}{\footnotesize{}}%
\begin{tabular}{@{}cc@{}}
\toprule 
{\hphantom{}Parameter} & {\hphantom{}Description}\\
\colrule 
{\hphantom{}$S/M,\,V/R$} & {\hphantom{}$S/M$ represents single or multiple launch(es) and
$V/R$ represents vectrize or replicate the kernel}\\
{\hphantom{}$N_{paral}$} & {\hphantom{}Value of the parellelization factor}\\
{\hphantom{}$N_{MultipleHP-H-preld}$} & {\hphantom{}preloaded data size of the }\textsc{\hphantom{}MultipleHP-H}{\hphantom{}
method }\\
{\hphantom{}$N_{MultipleHP-N-col}$} & {\hphantom{}Number of processed columns of all $N_{hp}$ harmonic
planes per work-group using }\textsc{\hphantom{}MultipleHP-N}\\
{\hphantom{}$N_{MultipleHP-R-col}$} & {\hphantom{}Number of processed columns of all $N_{hp}$ harmonic
planes per work-group using }\textsc{\hphantom{}MultipleHP-R}\\
{\hphantom{}$N_{points/wi}$} & {\hphantom{}Number of processed points of all $N_{hp}$ harmonic
planes per work item}\\
\botrule 
\end{tabular}
\end{wstable}

The \textsc{SingleHP} method is a straight-forward implementation
of Algorithm~\ref{alg:General-Harmonic-summing-Algorit}. The main
advantage of \textsc{MultipleHP} over \textsc{SingleHP} is that it
is unnecessary to store harmonic planes in the off-chip memory during
processing. The \textsc{MultipleHP-H }method is based on the Na\text{\"i}ve\textsc{-MultipleHP
}method, and it preloads the $N_{MultipleHP-H-preld}$ points with
the highest touching frequency in the FOP. Another loading method
is \textsc{MultipleHP-N} that loads all necessary points in the FOP
that are needed to calculate $N_{MultipleHP-N-col}$ columns of all
$N_{hp}$ harmonic planes. Though these methods can reduce the off-chip
memory accesses to some degree, the accesses to the off-chip memory
are still irregular. 

The \textsc{MultipleHP-R} method is based on the \textsc{MultipleHP-}N
method, however, the FOP is reordered and padded to generate the \textsc{r}FOP
before processing. In the \textsc{r}FOP, the necessary points in the
FOP that are needed to calculate a block of points in all $N_{hp}$
planes are stored in consecutive memory addresses. This makes some
points in the original FOP have to be stored in several places in
the \textsc{r}FOP, which leads to an increase in the \textsc{r}FOP
size. After reordering, the points in the \textsc{r}FOP can be streamed
to FPGA during processing. Besides $N_{MultipleHP-R-col}$, the parameter
$N_{points/wi}$ is an important factor for the \textsc{MultipleHP-}R
method, and it is restricted by the resources of the target FPGA. 

\begin{figure}
\begin{centering}
\includegraphics[bb=0bp 0bp 550bp 288bp,clip,scale=0.7]{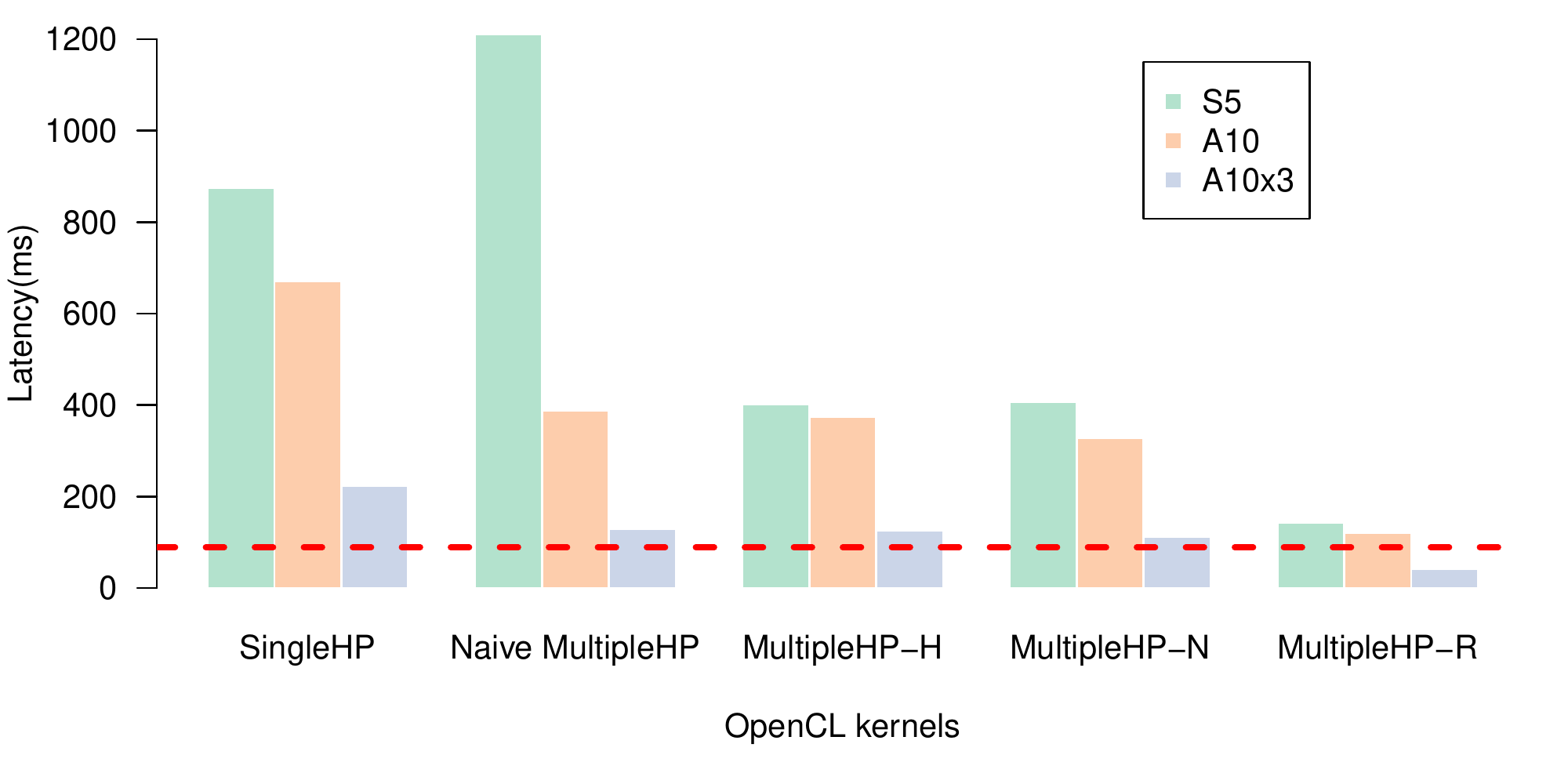}
\par\end{centering}
\caption{\label{fig:Performance-of-the}Execution latency of the straight-forward
and optimised methods on two types of FPGA devices (x3 means three
A10 FPGAs were used)}
\end{figure}

These different approaches were implemented using Intel FPGA-based
OpenCL. For each method, the parameters for the best performing implementation
and the resource usage and kernel execution latencies, including the
candidate detection part, are presented in Table~\ref{tab:Re-usage-HM}
and Figure~\ref{fig:Performance-of-the}. The red dot line in Figure~\ref{fig:Performance-of-the}
is the required time limitation and the execution latencies are for
processing half FOP. Kernel \textsc{MultipleHP-R} performs better
than the other kernels, however, additional processing has to be done
to reorder the standard FOP. The reorder task can be done on both
host and device. In the host program, the memory copy function~\texttt{memcpy()}
can handle this task efficiently. For the OpenCL kernel, there are
no such functions and a block of points has to be copied to another
place using a\texttt{ for} loop. Although the host processor has the
advantage over the FPGAs in reordering the FOP, the penalty for transferring
data between host and device has to be considered.
device has to be considered.

\begin{wstable}[h]
\caption{\label{tab:Re-usage-HM}Resource usage and execution latency of the
best harmonic-summing kernels on $A10$}
\begin{centering}
{\scriptsize{}}
\par\end{centering}{\scriptsize \par}
\centering{}{\scriptsize{}}%
\begin{tabular}{@{}ccccc@{}}
\toprule 
{\hphantom{}Kernel-(Settings)} & {\hphantom{}Logic utilization} & {\hphantom{}RAM blocks} & {\hphantom{}DSP Blocks} & {\hphantom{}Kernel frequency}\tabularnewline
 & {\hphantom{}} & {\hphantom{}} & {\hphantom{}} & {\hphantom{}$(MHz)$}\tabularnewline
\colrule 
\textsc{\hphantom{}SingleHP}{\hphantom{}-($S,R,8$)} & {\hphantom{}42\%} & {\hphantom{}28\%} & {\hphantom{}<1\%} & {\hphantom{}206.1}\\
{\hphantom{}Na\text{\"i}ve}\textsc{\hphantom{} MultipleHP} & {\hphantom{}17\%} & {\hphantom{}27\%} & {\hphantom{}<1\%} & {\hphantom{}227.06}\\
\textsc{\hphantom{}MultipleHP}{\hphantom{}-H-$(5\times2^{13})$} & {\hphantom{}19\%} & {\hphantom{}37\%} & {\hphantom{}<1\%} & {\hphantom{}235.84}\\
\textsc{\hphantom{}MultipleHP}{\hphantom{}-N-$(1)$} & {\hphantom{}17\%} & {\hphantom{}19\%} & {\hphantom{}<1\%} & {\hphantom{}276.54}\\
\textsc{\hphantom{}MultipleHP}{\hphantom{}-R-$(16,4)$} & {\hphantom{}30\%} & {\hphantom{}37\% } & {\hphantom{}3\%} & {\hphantom{}196.88}\\
\botrule 
\end{tabular}{\scriptsize \par}
\end{wstable}

Without including the candidate detection in the compilation and synthesis
process, \textsc{SingleHP-$(M,R,16)$} and \textsc{MultipleHP-R-$(64,8)$
}can be successfully synthesized for the $A10$ FPGA, however including
candidate detection (as done with the kernels in Figure~\ref{fig:Performance-of-the})
makes the synthesis fail due to exhausted FPGA resources. This is
the same type of compromise we will see later when the modules are
combined in the pipeline on the FPGA.

\section{\label{sec:OpenCL-based-Architecture}Combining Modules and Optimisation}

\subsection{\label{subsec:Design-Goals}Design Goals}

The FT convolution module and harmonic summing module are well-optimised,
and each can meet the required time limitation using three
high-end FPGAs. The goal of this research is to combine these two
modules while meeting the requirements of the FDAS module, especially
the time limitation. 

As introduced above, the FDAS module contains two main parts: 1) FT
convolution and 2) harmonic summing (including candidate detection).
The accumulated latency of different parts is the overall latency
$t_{FDAS}$ of the FDAS module in processing one input array, namely
the latencies of: the FT convolution (multiple FIR filters and power
calculation) $t_{FT}$, the FOP preparation $t_{FOP}$, and the harmonic
summing $t_{HM}$: 
\begin{equation}
t_{FDAS}=t_{FT}+t_{FOP}+t_{HM}.\label{eq:t_FDAS}
\end{equation}

Latency $t_{FT}$ is affected by three factors: the kernel launching
overhead $t_{klo}$, the execution latency of each FT convolution
kernel $t_{FT_{i}}^{'}$, and the number of times $N_{FT-launch}$
the kernel is launched. Hence, $t_{FT}$ can be expressed as 

\[
t_{FT}=\sum_{i=1}^{N_{FT-launch}}t_{FT_{i}}^{'}+\sum_{i=1}^{N_{FT-launch}}t_{klo_{i}}.
\]
Depending on the combination of the FDAS sub-modules, $t_{FOP}$
might consist of several parts such as discard $t_{discard}$, transpose
$t_{transpose}$, and reorder $t_{reorder}$ and can be expressed
as 
\[
t_{FOP}=\boldsymbol{B_{1}}t_{discard}+\boldsymbol{B_{2}}t_{transpose}+\boldsymbol{B_{3}}t_{reorder},
\]
 where the data types of $\boldsymbol{B_{1},\,B_{2},}$ and $\boldsymbol{B_{3}}$
are Boolean and the values depend on the combined sub-module kernels.
Latency $t_{HM}$ varies based on the applied method.

Regarding the FOP preparation, it is a module that is added between
the FT convolution module and the harmonic-summing module, which is
discussed in Section~\ref{subsec:FOP-Preparation}.

Based on the fastest results in Section~\ref{subsec:FT-Convolution-Module}
and Section~\ref{subsec:Harmonic-summing-Module}, even the achievable
$t_{FDAS}$ is greater than $t_{limit}$. Because of the limited logic
resources on the FPGA, the fastest implementations of two modules
cannot be merged into one implementation. There are two alternatives:
1) keep the optimised kernels and 2) modify the optimised kernels
to put the whole FDAS module in one FPGA device.

There are two options without modifying the optimised implementations:
1) use multiple FPGA devices or 2) reconfigure the FPGA device
several times. For the first solution, the data transfer rate between
the host and devices becomes an essential factor. With PCIe Gen3.0
for example, the theoretical latency of loading half FOP ($42\times2^{21}$
points) from one device and sending it to another device is about
the same as $t_{limit}$. If the FOP preparation module is assigned
to the host processor, it makes the overall pulsar search pipeline
impossible to meet the required time limit. Regarding the second solution,
it takes over one second for both $S5$ and $A10$ to reconfigure
the new bitstream file that is over 10x times larger than $t_{limit}$
which leaves alternative 2. If the optimised kernels are modified
to make all three modules fit into one FPGA device, $t_{FDAS}$ becomes
unimportant. The three parts of the FDAS module can work in parallel
in a pipeline by employing multiple buffering. Taking the triple buffering
as an example, each part can process points from different input arrays
at the same time, and the slowest section of these three kernels determines
the execution latency of a new input array. The combination of these
three parts becomes an important issue. In this research, we investigate
the suitable combination of the optimised implementations for a given
FPGA device. The total number of combinations is the product of the
number of FT convolution methods and the number of harmonic-summing
methods. These combinations can be categorised into four types: TDFIR
+ \textsc{SingleHP}, TDFIR + \textsc{MultipleHP,} FDFIR + \textsc{SingleHP}
and FDFIR + \textsc{MultipleHP.}

\subsection{\label{subsec:FOP-Preparation}FOP Preparation}

\begin{figure}
\begin{centering}
\includegraphics[bb=0bp 25bp 291bp 275bp,clip,scale=1]{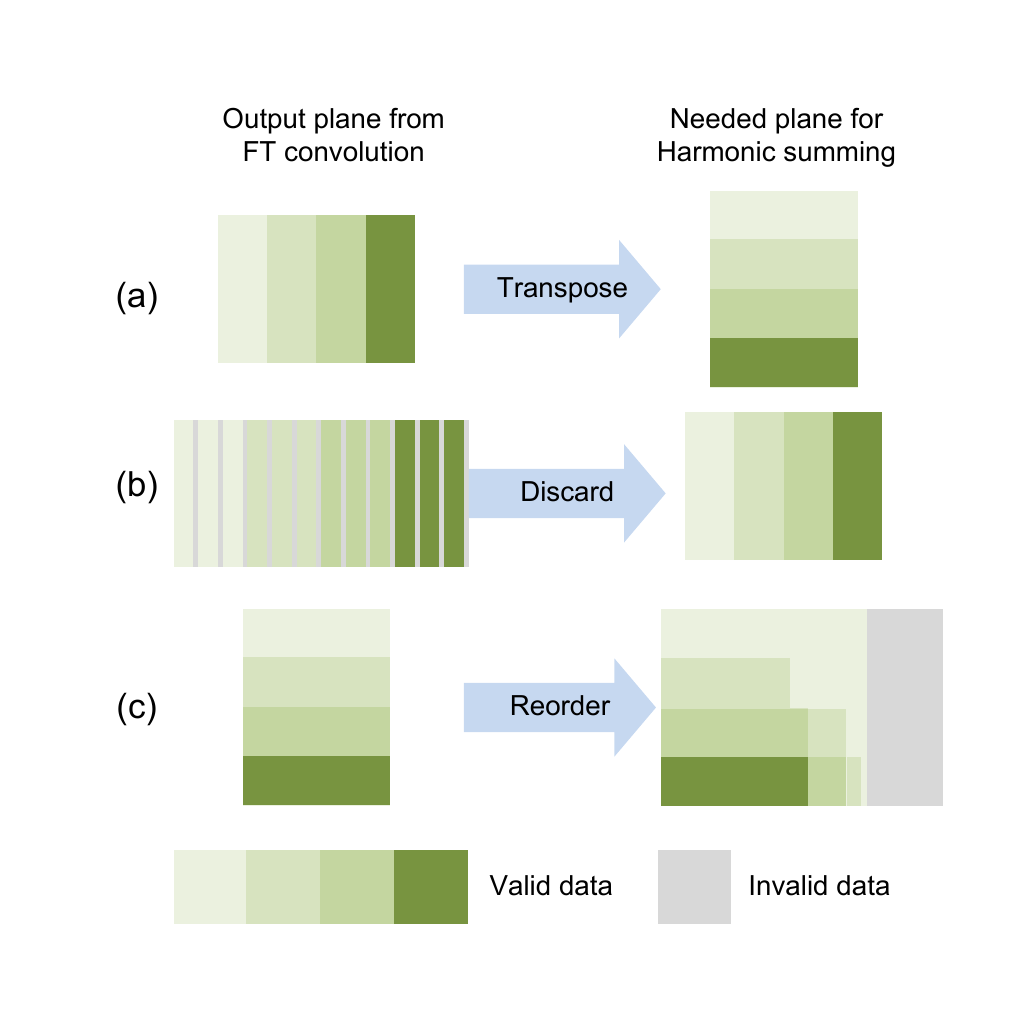}
\par\end{centering}
\caption{\label{fig:Three-types-of}Three types of processing the output plane
of the FT convolution module: (a) Transpose (b) Discard (c) Reorder.}

\end{figure}

As introduced in Section~\ref{subsec:FT-Convolution-Module} and
\ref{subsec:Harmonic-summing-Module}, the output plane from the FT
convolution and the needed plane for the harmonic summing varies based
on different kernel approaches. To make the FT convolution output
plane compatible with the harmonic summing input plane, the output
from the FT convolution module has to be transformed, and we add an FOP
preparation module to connect these two modules. There are three types
of transform processing: (a)~transpose, (b)~discard, and (c)~reorder,
which are depicted in Figure~\ref{fig:Three-types-of}.

For the TDFIR-based FT convolution kernels, each row of the output
plane is the output from an FIR filter (Figure~\ref{fig:Three-types-of}(a)).
However, processing column by column might be more efficient for some
harmonic summing kernels. In this case, the output plane has to be
transposed. 

The output plane of the FDFIR-based FT convolution kernels (Figure~\ref{fig:Three-types-of}(b))
contains a number of slices of dummy/invalid points and these points
need to be discarded to get the standard FOP.

The \textsc{MultipleHP}-R kernel performs better than other \textsc{MultipleHP}-based
harmonic-summing kernels, however, the input plane is not the standard
FOP but the reordered FOP (\textsc{r}FOP). To generate the \textsc{r}FOP,
the output plane has to be padded and reordered (Figure~\ref{fig:Three-types-of}(c)).
The reason for padding with dummy data is to make the number of loaded
points per clock cycle a power of two, which is more efficient than
other numbers.

For different kernel combinations, these three types of transforms
can be combined. If the output plane is the same as the needed input
plane, the FOP preparation module can be removed. For example, if
the FT convolution output plane is the left plane in Figure~\ref{fig:Three-types-of}(b)
and the needed input is the right plane in Figure~\ref{fig:Three-types-of}(c),
all these three transforms have to be applied in a certain order (discard
+ transpose + reorder) in the FOP preparation kernel.

\subsection{\label{subsec:Combining-Kernels-(Optimizing}Pipeline Computing }

\begin{figure}
\begin{centering}
\includegraphics[bb=30bp 20bp 520bp 330bp,clip,scale=0.8]{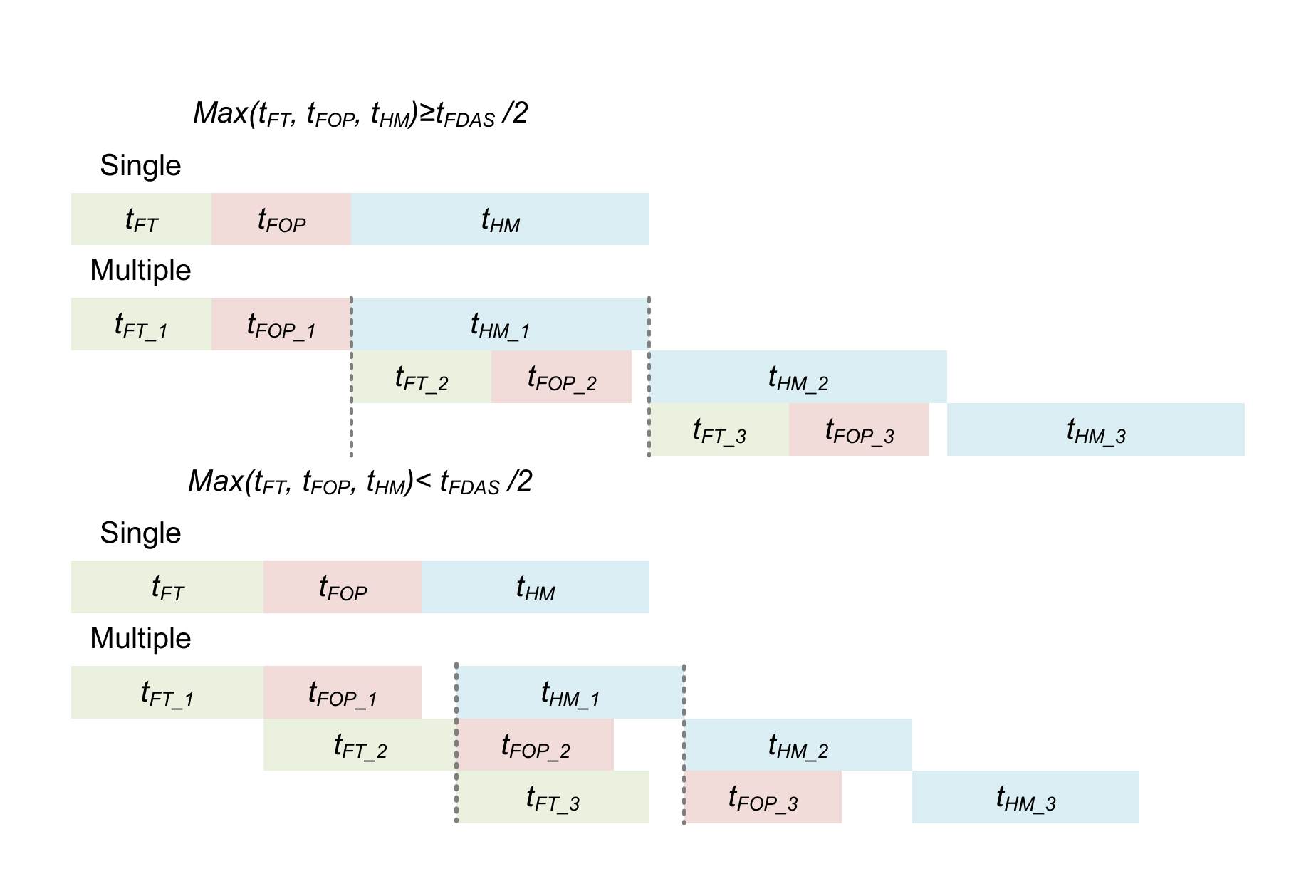}
\par\end{centering}
\caption{\label{fig:buffer_type}Execution latency of single and multiple input
arrays using double and triple buffering }
\end{figure}

Instead of processing one input array, the FDAS module keeps running
(24/7/365) when it is employed and will process a constant stream
of input signals. The main purpose of this research is to optimise
the execution latency of multiple input arrays, i.e. the throughput,
but not the overall execution latency of a single input array $t_{FDAS}$.
Therefore, we investigate the pipeline processing of the FDAS module.
Given the three sub-modules, the ideal execution latency for each
input array in a pipeline, which is the pipeline period, is $max(t_{FT},\,t_{FOP},\,t_{HM})$
and the number of required buffers depends on $t_{FDAS}$ and $max(t_{FT},\,t_{FOP},\,t_{HM})$,
which is illustrated in Figure~\ref{fig:buffer_type}. If $max(t_{FT},\,t_{FOP},\,t_{HM})\geq t_{FDAS}/2$,
double buffering can be employed, and when $max(t_{FT},\,t_{FOP},\,t_{HM})<t_{FDAS}/2$,
it is recommended to adopt triple buffering. 

Note that $t_{FDAS}$ of the two combinations in Figure~\ref{fig:buffer_type}
are the same, but the $max(t_{FT},\,t_{FOP},\,t_{HM})<t_{FDAS}/2$
combination performs better than the $max(t_{FT},\,t_{FOP},\,t_{HM})\geq t_{FDAS}/2$
combination when employing pipeline processing, as the pipeline stages
are more balanced in the latter case. For combinations where $max(t_{FT},\,t_{FOP},\,t_{HM})\geq t_{FDAS}/2$,
the parallelisation factors of the three combined kernels can be adjusted
to reduce $max(t_{FT},\,t_{FOP},\,t_{HM})$ to a value smaller than
$t_{FDAS}/2$ while aiming that $t_{FDAS}$ is not increased. In other
words, the objective of our research here is to minimise $max(t_{FT},\,t_{FOP},\,t_{HM})$
within the resource and bandwidth limits of the FPGA by carefully
investigating how to best combine and configure the optimised kernels
of the sub-modules.

\subsubsection{Device Limitation}

For most of the accelerators, the FPGA devices are connected to the
host processor through the PCIe bus (Figure~\ref{fig:FPGA-devices-as}),
which is introduced in Section~\ref{sec:FDAS-Module}. Three major
parts can limit the device performance: FPGA resources, off-chip memory,
and the data transfer bus.

\paragraph{FPGA Resources}

The logic cells, DSP blocks, and (embedded) RAM blocks are three main
types of FPGA resources, and the limit of each kind of resource leads
to different problems. The logic cells are employed to handle the
necessary fixed-point operations and the shift register. The number
of DSP blocks decides the number of parallelised floating-point operations
such as multiplications. Regarding the RAM blocks, they are the main
on-chip memory, and the number of RAM blocks restricts the amount
of data that can be stored in local memory during processing.

\paragraph{Off-chip Memory}

Two factors regarding the off-chip memory are discussed: 1) data transfer
rate and 2) off-chip memory size.

Because the FOP is stored in off-chip memory, the transfer rate between
FPGA and off-chip memory affects the overall performance directly.
The off-chip memory type and the width of the connected data bus are
factors that determine the theoretical transfer rate. The FPGA acceleration
cards employed in this research use DDR3 memory and the $HARP$ platform
uses DDR4 memory. Regarding the bit-width of the data bus, the $S5$
card has two memory banks and connects each memory bank with a 64-bit
data bus, which has 128-bit data bus in total. Regarding the $A10$
card, each memory bank is connected with a 72-bit data bus, and the
sum of two memory banks is 144. Hence, under the same operation frequency,
the data transfer rate of $A10$ is higher than that of $S5$.

The off-chip memory size affects the performance especially when multiple
buffering is adopted. Take the triple buffering (in Figure~\ref{fig:buffer_type})
as an example, if the off-chip memory size is not large enough to
hold three FOPs but will hold two FOPs, the implementation is restricted to
double buffering. In this case, the execution latency for a new input
array might be increased to $t_{FDAS}-max(t_{FT},\,t_{FOP},\,t_{HM})$,
which is larger than $t_{FDAS}/2$, assuming that $max(t_{FT},\,t_{FOP},\,t_{HM})\leq t_{FDAS}/2$
(the case for triple buffering).

\paragraph{Data Transfer Bus}

The PCIe bus is the main connection between the host processor and
FPGA devices. The transfer rate affects the performance especially
when the data has to be transferred between the host and the device during
processing. It is determined by the generation of the PCIe bus and
the number of lanes connected to the FPGA devices. For example, PCIe
Gen3.0 (used in the $A10$ board) provides $8.0GTransfers/s$ per
lane, while the latest Gen4.0 provides $16.0GTransfers/s$ per lane.
The number of lanes can vary between 1 and 16 but is usually either
8 (used in the $S5$ and $A10$ board) or 16 for FPGA acceleration
cards.

Besides the PCIe bus, the Intel QuickPath Interconnect (QPI) is employed
in $HARP$. It is a point-to-point interconnect released by Intel.
The QPI can be operated at up to $4.8GHz$ and the data transfer rate
can be tens of $GBytes/s$.

\subsubsection{Performance Factors}

The performance of the pipelined FDAS module is mainly influenced
by three factors: 1) parallelisation factor for each sub-module, 2)
maximum frequency of the kernels, and 3) the global memory bandwidth. 

\paragraph{Parallelisation Factor }

The optimised kernels as discuss in Section~\ref{sec:FDAS-Module}
almost fully exploit the target devices (such as their logic resources
and off-chip memory bandwidth) and some kernels completely exhaust
one type of resource such as the TDFIR kernel on $S5$ consumes all DSP
blocks. To integrate several kernels on one FPGA device, the optimised
kernels need to compromise each other, and the straight-forward
solution is to reduce the parallelisation factors of the optimised
kernel. This obviously leads to an increase of the execution latency
of the individual kernels.

\paragraph{Kernel Frequency}

The high percentage of resource usage of a combined kernel makes it
complex and hard to be implemented by the OpenCL compiler and synthesis
tools. This affects the maximum kernel frequency at which it can run,
which directly influences the performance. 

\paragraph{Off-chip Memory Bandwidth}

In pipeline computing, two or three kernels are executed simultaneously
(Figure~\ref{fig:buffer_type}). If the total needed off-chip memory
bandwidth surpasses the theoretical off-chip memory bandwidth, these
kernels might not perform as fast as when executed. In this case,
the maximum execution latency $max(t_{FT},\,t_{FOP},\,t_{HM})$ is
increased and the performance drops. 

\subsection{\label{subsec:Host-and-Device}Host and Device}

\subsubsection{Data Transfer Approaches }

For FPGA-based OpenCL, there are mainly two types of data transfer
approaches between host and accelerator (FPGA). 1) general buffer
transfer and 2) shared virtual memory (SVM). 

\paragraph{General Buffer Transfer}

In an OpenCL host program, a buffer object (one-dimensional) can be
transferred between the device off-chip memory (i.e., OpenCL global memory)
and the host memory using the \texttt{clEnqueueReadBuffer} and\texttt{
clEnqueueWriteBuffer} functions. For two- or three-dimensional buffer,
the \texttt{clEnqueueReadImage} and\texttt{ clEnqueueWriteImage} are
employed. The transfer is realised via the PCIe bus and the rate depends
on its specification, see above.

\paragraph{Shared Virtual Memory }

\begin{figure}
\begin{centering}
\includegraphics[scale=0.8]{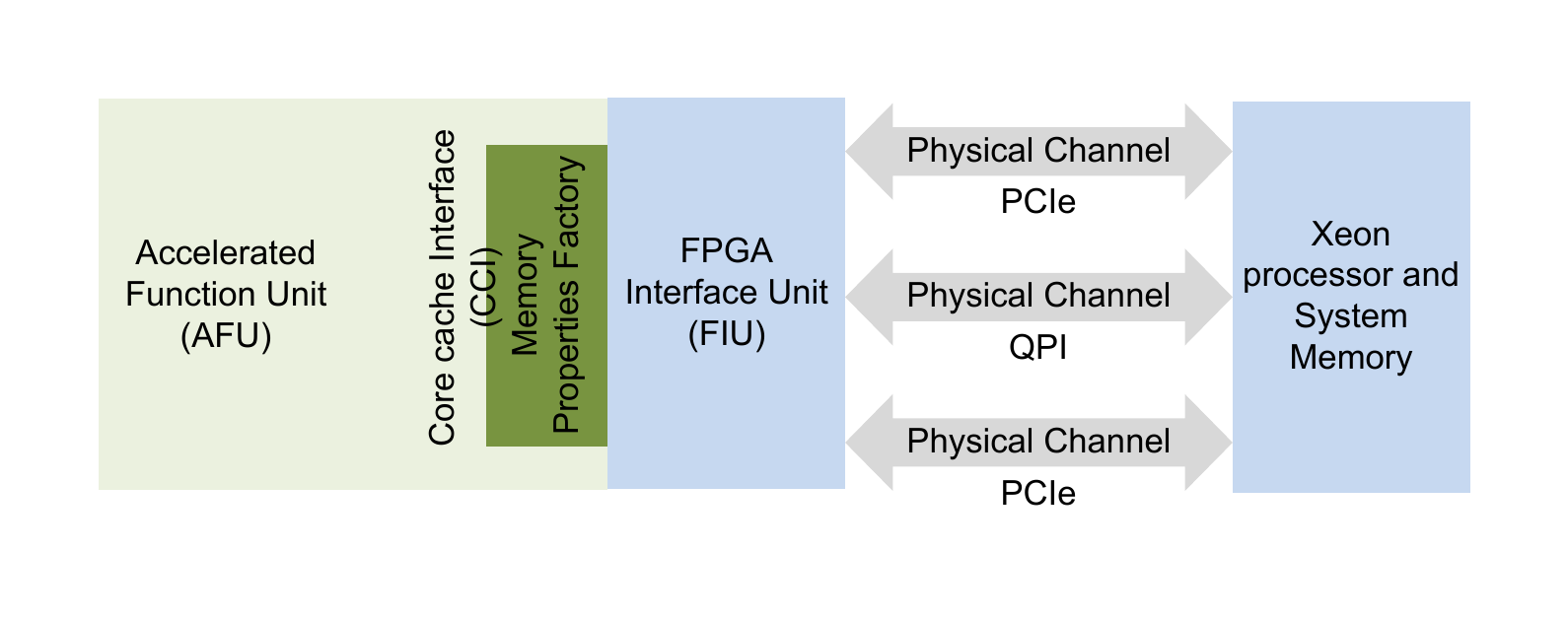}
\par\end{centering}
\caption{\label{fig:xeon-fpga}Intel Xeon Processor with FPGA IP}
\end{figure}

Using shared virtual memory (SVM) is a technique to extend the (OpenCL)
global memory region into the host memory region. It is supported
by the OpenCL 2.0 specification, and the host processor and device(s)
need a shared memory system. Since the $A10$ and $S5$ devices have
no physical shared memory with the host and the SVM technique is
not supported there. For the Intel Xeon processor platform with an
integrated Arria 10 FPGA, referred to as Xeon+FPGA, the FPGA and processor
are in the same package. An illustration of this is given is Figure~\ref{fig:xeon-fpga}.
Inside the FPGA, the accelerated function unit (AFU) is available
to be programmed by the developer, the other interfacing blocks are
provided by Intel. The core cache interface (CCI) provides a base
platform memory interface that exposes physical channels as a single,
multiplexed read/write memory interface. The embedded FPGA is connected
to the computer system memory (DDR4) through several physical channels
such as PCIe and Intel QPI.  

The memory properties factory (MPF) block is optional. When it is
employed, it is instantiated as a CCI-to-CCI bridge, maintaining the
same interface but adding new semantics. The main advantage of MPF
is that it can translate the virtual address to a physical address and
the FPGA and CPU can share pointers with each other. 

SVM-based transfer is about 2x times faster than that of the general
buffer-based transfer. By adding the FPGA to the chip-package, the
physical design has to compromise with many additional constraints,
and the performance of the processor part might not provide the same
performance as the independent package processors of the same technology.

\subsubsection{Tasks on the host}

The FPGA devices are employed as the accelerator, so naturally distributing
and balancing tasks between the host and the device are investigated.
Due to the usual performance penalty for transferring data between
host and device, it is recommended to execute most or all tasks on
the device. However, there are situations where data has to be transferred
back to the host during processing.

1) The execution latency of a task on the host is significantly faster
than that on the device.

Although FPGA devices perform better than the host in a wide range
of applications, they still have a weak point in serial processing.
If a task is arranged to be executed by the host, the data transfer
rate between the host and devices becomes the main issue. Hence, when
determining the performance advantage of the host processor or FPGAs,
the inflicted data transfer delay needs to be considered in the analysis.

2) Data dependency of using multiple devices

This situation happens when multiple devices are employed in executing
the same input array. The host can then become the master that needs
to manage the dependences and communications between the sub-tasks
and the devices. Of course, the ideal case in designing the FDAS module
is to avoid transferring data between the host and devices while processing
as much as possible.

\subsection{\label{subsec:Multiple-Devices}Multiple Devices}

When employing more than one FPGA device for the acceleration, there
are two obvious approaches: using 1) the same configuration (bitstream)
file (single) or 2) different configuration files (multiple) for the
programming of the FPGA devices.

\subsubsection{Single Configuration File}

\paragraph{Single Input Array}

Multiple devices for a single input array can be necessary if $t_{FDAS}>t_{limit}$.
Except for the \textsc{MultipleHP-R} method in Section~\ref{subsec:Harmonic-summing-Module},
the optimised harmonic-summing implementations on a single device
take longer than the required time limit. When multiple devices ($N_{devices}$)
are employed, the harmonic-summing task can be split into $N_{devices}$
independent parts and each FPGA device processes $1/N_{devices}$
of the FOP. In this case, the ideal execution latency drops to $max(t_{FT},\,t_{FOP},\,t_{HM}/N_{devices})$.

For the FT convolution module and FOP preparation module, each of
the $N_{devices}$ devices generates the full FOP, and it is unnecessary
for a device to communicate with other devices while processing. Processing
a single input array while all devices are configured with the same
bitstream file, the same FOP is generated $N_{devices}$ times. 

\paragraph{Multiple Input Arrays}

For multiple input arrays, the host sends $N_{devices}$ different
input array to $N_{devices}$ FPGA devices and $N_{devices}$ input
arrays are processed in parallel. Compared with a single device,
the ideal execution latency for multiple devices reduces to $max(t_{FT},\,t_{FOP},\,t_{HM})/N_{devices}$, 

\[
\frac{max(t_{FT},\,t_{FOP},\,t_{HM})}{N_{devices}}\leq max(t_{FT},\,t_{FOP},\,\frac{t_{HM}}{N_{devices}}).
\]
Hence, the multiple input arrays approach has a theoretical advantage
when $t_{HM}=max(t_{FT},\,t_{FOP},\,t_{HM})$.

\subsubsection{Multiple Configuration Files}

For some combinations, the FT convolution and harmonic-summing have
to compromise with each other by reducing their parallelisation factors
or scales. This leads to a decrease in performance for both parts.
By using multiple devices, each device can be configured with one
or two functions while taking full advantage of the device resources.
In Figure~\ref{fig:buffer_type}, each stage can be assigned to a
device and the number of buffering equals the number of devices.
For example, when $max(t_{FT},\,t_{FOP},\,t_{HM})<t_{FDAS}/2$, three
devices need to be installed and each device is configured with only
one module.

The main problem with this method is the frequent communication between
the host and the devices. The host needs to keep organizing data between
different devices. This method requires a high transfer rate between
the host and devices such as high generation PCIe and QPI.

\subsection{A Case Study\label{subsec:A-Case-Study}}

\begin{figure}
\begin{centering}
\includegraphics[bb=20bp 25bp 370bp 280bp,clip,scale=1]{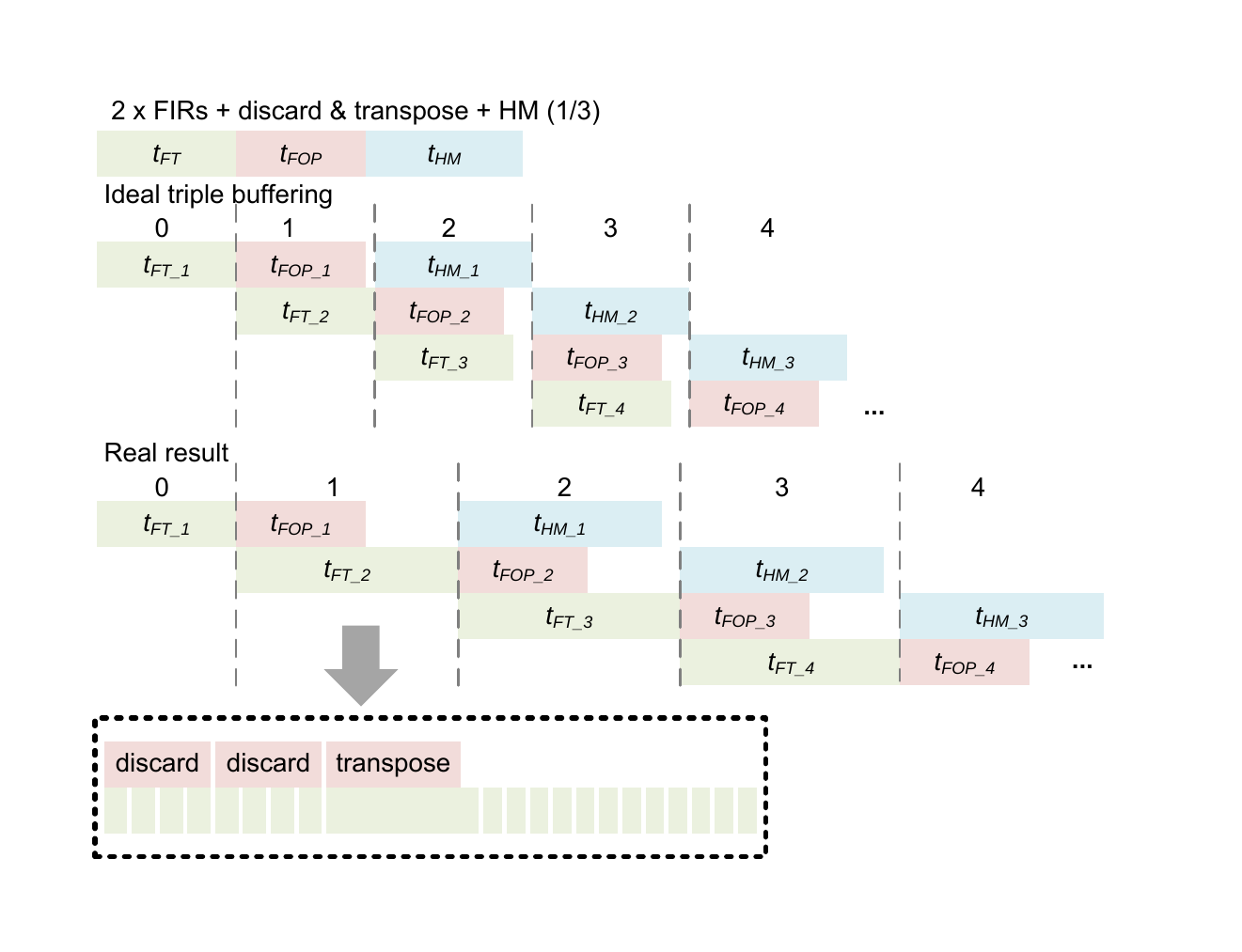}
\par\end{centering}
\caption{\label{fig:The-limitation-of}Ideal and real latencies of the kernels
of the case study FDFIR+\textsc{MultipleHP-N}}
\end{figure}

Before we systematically evaluate the pipeline design and the many
combinations of the different sub-module kernels, let us have a closer
look at the combination of the FDFIR+\textsc{MultipleHP-N} kernels
as a case study. The execution latency of one input array using three
devices is depicted in Figure~\ref{fig:The-limitation-of} (top).
Three devices are configured with the same file and the harmonic-summing
part of each device processes $1/3$ of a half-FOP. The FDFIR filter
is parallelised twice (i.e., two filters working in parallel), so
the FT convolution kernel needs to be launched $N_{FT-launch}=21$
times (as there are 42 filters to be applied). Ignoring the FT convolution
kernel launching overhead, the execution latency $t_{FT}$ is $\sum_{i=1}^{21}t_{FT_{i}}^{'}$.
The FOP preparation kernel contains discard and transpose, and the
harmonic summing kernel processes $1/3$ of the overall task. 

As can be seen in the basic execution latency (top), $t_{HM}\geq max(t_{FT},\,t_{FOP})$
but smaller than $t_{FDAS}/2$. Based on the discussion in Section~\ref{subsec:Combining-Kernels-(Optimizing},
we can infer the execution latency of multiple input arrays using
triple buffering. Ideally, the execution latency of each part remains
the same as that of executing one input array. The time cost for one
new input array is $t_{HM}$ in this example, Figure~\ref{fig:The-limitation-of}
(Middle). 

However, the real execution latency of multiple input arrays takes
much longer than the ideal case. The real result and details are given
in Figure~\ref{fig:The-limitation-of} (Bottom) as well. When the
FOP preparation part is processing, the FT convolution part is severely
affected, and the $t_{HM}$ is increased as well. Because two FIR
filters are working in parallel, the discard kernel is launched twice
for two output groups. In the zoomed in part, during the discard processing,
the FT convolution kernels are launched 8x times to process the next
input array using 16 FIR filters, and the $9th$ FT convolution kernel
is launched with the transpose kernel. The average value of $t_{FT_{1}}^{'}$
to $t_{FT_{8}}^{'}$is larger the that of $t_{FT_{10}}^{'}$to $t_{FT_{21}}^{'}$
and $t_{FT_{9}}^{'}$ is several times larger than others. The value
of $t_{FT_{9}}^{'}$ is about the same as $t_{discard}+\sum_{i=9}^{21}t_{FT_{i}}^{'}/12$. 

The main reason for the stretch of $t_{FT_{1}}^{'}$to $t_{FT_{9}}^{'}$is
the limited global memory bandwidth (GMB) of the FPGA device. The
discard, transpose, and FT convolution kernels all depend heavily
on the GMB. When two of them are processing at the same time, the
needed GMB exceeds the device GMB. For the transpose kernel, it exhausts
the device GMB while processing alone. When the transpose and FT convolution
kernels are launched together, the FT convolution kernel is pended
until the transpose kernel has been finished. When the three parts
are processing in parallel, the value of $t_{FT}$ will be larger
than that of in the zoomed part in Figure~\ref{fig:The-limitation-of}.
In real processing, the FT convolution becomes the dominant kernel
and it determines the time delay until the next new input array can
be processed (i.e., the pipeline period).

\section{\label{sec:Evaluation}Experimental Evaluation}

This section experimentally evaluates the design space of the FDAS
module pipeline, by considering a large number of combinations of
the optimised kernels of the sub-modules and their design parameters.
The advantage of the multiple buffering technique is evaluated and
multiple acceleration devices are employed to accelerate the FDAS
module. The combinations are assessed according to their resource
usage, execution latency, and energy dissipation and power consumption.

\subsection{Resource Usage}

High-end Arria 10 FPGAs (Nallatech 385A with Intel Arria 10 GX1150,
in Table~\ref{tab:Details-of-FPGA}) are employed for the experiments.
All combinations are implemented using Intel FPGA-based OpenCL, and
all combined FDAS kernels are compiled using AOCL version 16.0.0.222. 

For the FT convolution module, the OLA-TD and OLS-FD methods, in Section~\ref{subsec:FT-Convolution-Module},
are used. The OLA-$N_{paral}$ kernel, which parallelises $N_{paral}$
complex SPF multiplications, and the AOLS-$N_{OLS-FT}$, which split
the input array into a group of chunks whose each length is $N_{OLS-FT}$,
are employed to combine with harmonic-summing modules. Taking the OLA-128
kernel as an example, it has to be launched four times to implement
a 421-tap FIR filter, and its execution latency is the same
as would be for a 512-tap FIR filter, as 91 taps are unused (set to
zero). 

For the harmonic summing modules, the \textsc{SingleHP-$(S,R,8)$
}kernel is selected for the \textsc{SingleHP} method and the Na\text{\"i}ve\textsc{-MultipleHP,}
\textsc{MultipleHP-N-$(1)$, }and\textsc{ MultipleHP-R-$(16,4)$}
are chosen for the \textsc{MultipleHP} method. The parallelisation
factors of the harmonic-summing module are the largest values that
can be compiled successfully by the AOC when combining with the FT
convolution module. The \textsc{MultipleHP-H} method is based on the
Na\text{\"i}ve\textsc{-MultipleHP }method and the best-performing
implementation, which is \textsc{MultipleHP-H}-$(5\times2^{13})$,
cannot be combined with other kernels as it exhausts the available
FPGA resources. When the $N_{MultipleHP-H-preld}$ is decreased to
reduce resources, it performs worse than the Na\text{\"i}ve\textsc{-MultipleHP}
kernel, so it is not considered in this research. 

In summary, for the two types of FT convolution and the four types
of the harmonic-summing, there are a total of eight FDAS combinations,
listed in detail in Table~\ref{tab:Resource-usage-of}. The table
also provides resource usage and achievable frequencies of these combinations.
Because each of the FT convolution and harmonic-summing parts contains
at least one kernel, there are two or more independent kernels in
the FDAS module. Each kernel in the FDAS module is compiled as an
independent kernel. To arrange multiple independent kernels in a target
FPGA, the compiler has to add more restrictions than compiling a single
kernel. Although some successfully compiled kernels use less than
half of the device resources, the parallelisation factors still cannot
be increased as the compilation then fails. For each of the combinations
in Table~\ref{tab:Resource-usage-of}, several parallelisation factors
are tested, and the combination with the fastest execution latency
is recorded. Taking combination FDFIR+\textsc{MultipleHP-R} as an
example, AOLS-1024, AOLS-2048, and AOLS-4096 are combined with \textsc{MultipleHP-R-}\textsc{\small{}$(16,4)$,
}\textsc{MultipleHP-R-}\textsc{\small{}$(16,8)$, }\textsc{MultipleHP-R-}\textsc{\small{}$(64,4)$,}
and\textsc{\small{} }\textsc{MultipleHP-R-}\textsc{\small{}$(64,8)$,
}which are 12 combinations in total. Among these 12 combinations,
only three combinations can be successfully compiled. Of these three
combinations, AOLS-2048+\textsc{\small{}MultipleHP-R-$(16,4)$} provides
better performance than the other successfully compiled combinations
such as AOLS-1024+\textsc{MultipleHP-R-}\textsc{\small{}$(16,4)$},
hence it is the one recorded in the table\textsc{\small{}.}{\small \par}

\begin{wstable}[h]
\caption{\label{tab:Resource-usage-of}Resource usage of the combined FDAS
kernels}
\centering{}%
\begin{tabular}{@{}cccccccc@{}}
\toprule 
\multicolumn{2}{c}{{\small{}FT Convolution}} & {\small{}FOP preparation} & {\small{}Harmonic-summing} & {\small{}Frequency} & {\small{}Logic } & {\small{}DSP } & {\small{}RAM }\tabularnewline
\multicolumn{2}{c}{{\small{}module}} & {\small{}module} & {\small{}module} & {\small{}$(MHz)$} & {\small{}utilisations} & {\small{}blocks} & {\small{}blocks}\tabularnewline
\colrule  
\multirow{6}{*}{\rotatebox{90}{TDFIR}} & {\small{}OLA-128} & {\small{}\textendash{}} & {\small{}Na}\text{\"i}{\small{}ve-}\textsc{\small{}MultipleHP} & {\small{}207.8} & {\small{}25\%} & {\small{}44\%} & {\small{}42\%}\tabularnewline
 & {\small{}OLA-256} & {\small{}\textendash{}} & {\small{}Na}\text{\"i}{\small{}ve-}\textsc{\small{}MultipleHP} & {\small{}207.1} & {\small{}32\%} & {\small{}86\%} & {\small{}64\%}\tabularnewline
 & {\small{}OLA-128} & {\small{}transpose} & \textsc{\small{}MultipleHP-N-(1)} & {\small{}159.1} & {\small{}37\%} & {\small{}43\%} & {\small{}56\%}\tabularnewline
 & {\small{}OLA-128} & {\small{}transpose+reorder} & \textsc{\small{}MultipleHP-R-$(16,4)$} & {\small{}182.2} & {\small{}47\%} & {\small{}46\%} & {\small{}52\%}\tabularnewline
 & {\small{}OLA-128} & {\small{}\textendash{}} & \textsc{\small{}MultipleHP-R-$(16,4)$} & {\small{}171.9} & {\small{}37\%} & {\small{}46\%} & {\small{}40\%}\tabularnewline
 & {\small{}OLA-128} & {\small{}\textendash{}} & \textsc{\small{}SingleHP-$(S,R,8)$} & {\small{}179.6} & {\small{}45\%} & {\small{}44\%} & {\small{}49\%}\tabularnewline
\multirow{6}{*}{\rotatebox{90}{FDFIR}} & {\small{}AOLS-2048 } & {\small{}discard+transpose} & {\small{}Na}\text{\"i}{\small{}ve-}\textsc{\small{}MultipleHP} & {\small{}178.4} & {\small{}41\%} & {\small{}23\%} & {\small{}78\%}\tabularnewline
 & {\small{}AOLS-2048 } & {\small{}discard+transpose} & \textsc{\small{}MultipleHP-N-$(1)$} & {\small{}175} & {\small{}35\%} & {\small{}12\%} & {\small{}66\%}\tabularnewline
 & {\small{}AOLS-2048 } & {\small{}discard+transpose} & \multirow{2}{*}{\textsc{\small{}MultipleHP-R-$(16,4)$}} & \multirow{2}{*}{{\small{}180}} & \multirow{2}{*}{{\small{}49\%}} & \multirow{2}{*}{{\small{}15\%}} & \multirow{2}{*}{{\small{}56\%}}\tabularnewline
 &  & {\small{}+reorder} &  &  &  &  & \tabularnewline
 & {\small{}AOLS-2048 } & {\small{}\textendash{}} & \textsc{\small{}MultipleHP-R-$(16,4)$} & {\small{}185.2} & \multirow{1}{*}{{\small{}35\%}} & {\small{}15\%} & {\small{}42\%}\tabularnewline
 & {\small{}AOLS-2048 } & {\small{}discard} & \textsc{\small{}SingleHP-$(S,R,8)$} & {\small{}196.5} & {\small{}37\%} & {\small{}23\%} & {\small{}70\%}\tabularnewline
\botrule 
\end{tabular}
\end{wstable}

\begin{wstable}[h]
\caption{\label{fig:ru-indenpendnet-combined}Resource usage and kernel frequency
of independent and combined implementations}

{\footnotesize{}}{\footnotesize \par}
\centering{}{\footnotesize{}}%
\begin{tabular}{@{}ccccccccccccc@{}}
\toprule 
\multirow{2}{*}{{\footnotesize{}Kernels}} & \multicolumn{3}{c}{{\footnotesize{}Frequency ($HMz$)}} & \multicolumn{3}{c}{{\footnotesize{}Logic utilization}} & \multicolumn{3}{c}{{\footnotesize{}DSP blocks}} & \multicolumn{3}{c}{{\footnotesize{}RAM blocks}}\tabularnewline
 & {\footnotesize{}FT} & {\footnotesize{}HM} & {\footnotesize{}Comb.} & {\footnotesize{}FT} & {\footnotesize{}HM} & {\footnotesize{}Comb.} & {\footnotesize{}FT} & {\footnotesize{}HM} & {\footnotesize{}Comb.} & {\footnotesize{}FT} & {\footnotesize{}HM} & {\footnotesize{}Comb.}\tabularnewline
\colrule 
{\footnotesize{}OLA-128 +} & \multirow{2}{*}{{\footnotesize{}267.4}} & \multirow{2}{*}{{\footnotesize{}227.06}} & \multirow{2}{*}{{\footnotesize{}207.9}} & \multirow{2}{*}{{\footnotesize{}21\%}} & \multirow{2}{*}{{\footnotesize{}17\%}} & \multirow{2}{*}{{\footnotesize{}25\%}} & \multirow{2}{*}{{\footnotesize{}42\%}} & \multirow{2}{*}{{\footnotesize{}<1\%}} & \multirow{2}{*}{{\footnotesize{}44\%}} & \multirow{2}{*}{{\footnotesize{}27\%}} & \multirow{2}{*}{{\footnotesize{}27\%}} & \multirow{2}{*}{{\footnotesize{}42\%}}\tabularnewline
{\footnotesize{}Na}{\small{}\text{\"i}}{\footnotesize{}ve-}\textsc{\footnotesize{}MultipleHP} &  &  &  &  &  &  &  &  &  &  &  & \tabularnewline
{\footnotesize{}OLA-128 +} & \multirow{2}{*}{{\footnotesize{}267.4}} & \multirow{2}{*}{{\footnotesize{}206.1}} & \multirow{2}{*}{{\footnotesize{}179.66}} & \multirow{2}{*}{{\footnotesize{}21\%}} & \multirow{2}{*}{{\footnotesize{}42\%}} & \multirow{2}{*}{{\footnotesize{}45\%}} & \multirow{2}{*}{{\footnotesize{}42\%}} & \multirow{2}{*}{{\footnotesize{}<1\%}} & \multirow{2}{*}{{\footnotesize{}44\%}} & \multirow{2}{*}{{\footnotesize{}27\%}} & \multirow{2}{*}{{\footnotesize{}28\%}} & \multirow{2}{*}{{\footnotesize{}49\%}}\tabularnewline
\textsc{\footnotesize{}SingleHP-$(S,R,8)$} &  &  &  &  &  &  &  &  &  &  &  & \tabularnewline
{\footnotesize{}OLA-128 +} & \multirow{2}{*}{{\footnotesize{}267.4}} & \multirow{2}{*}{{\footnotesize{}196.9}} & \multirow{2}{*}{{\footnotesize{}171.9}} & \multirow{2}{*}{{\footnotesize{}21\%}} & \multirow{2}{*}{{\footnotesize{}30\%}} & \multirow{2}{*}{{\footnotesize{}37\%}} & \multirow{2}{*}{{\footnotesize{}42\%}} & \multirow{2}{*}{{\footnotesize{}3\%}} & \multirow{2}{*}{{\footnotesize{}46\%}} & \multirow{2}{*}{{\footnotesize{}27\%}} & \multirow{2}{*}{{\footnotesize{}37\%}} & \multirow{2}{*}{{\footnotesize{}40\%}}\tabularnewline
\textsc{\footnotesize{}MultipleHP-R-$(16,4)$} &  &  &  &  &  &  &  &  &  &  &  & \tabularnewline
{\footnotesize{}AOLS-2048+} & \multirow{2}{*}{{\footnotesize{}252.3}} & \multirow{2}{*}{{\footnotesize{}196.9}} & \multirow{2}{*}{{\footnotesize{}185.2}} & \multirow{2}{*}{{\footnotesize{}16\%}} & \multirow{2}{*}{{\footnotesize{}30\%}} & \multirow{2}{*}{{\footnotesize{}35\%}} & \multirow{2}{*}{{\footnotesize{}11\%}} & \multirow{2}{*}{{\footnotesize{}3\%}} & \multirow{2}{*}{{\footnotesize{}15\%}} & \multirow{2}{*}{{\footnotesize{}32\%}} & \multirow{2}{*}{{\footnotesize{}37\%}} & \multirow{2}{*}{{\footnotesize{}42\%}}\tabularnewline
\textsc{\footnotesize{}MultipleHP-R-$(16,4)$} &  &  &  &  &  &  &  &  &  &  &  & \tabularnewline
\botrule 
\end{tabular}{\footnotesize \par}
\end{wstable}

Among these combinations, the OLA-128+Na\text{\"i}ve-\textsc{MultipleHP}
and OLA-128+\textsc{SingleHP-$(S,R,8)$} do not require the FOP preparation
kernel. For combinations that contain the \textsc{MultipleHP-R-$(16,4)$}
kernel, if the FOP preparation task is assigned to the host processor,
the FOP preparation module does not need to be implemented in the
FPGA. The resource usage and kernel frequency of independent and combined
implementations of these four kernels are given in Table~\ref{fig:ru-indenpendnet-combined}.
As expected, the frequency of the combined kernel is lower than each
of its element kernels. The DSP block usage is slightly larger than
the sum of element kernels. The logic cell and RAM blocks utilisations
of a combined kernel are larger than that of each element kernel,
however, smaller than the sum of all element kernels. The reason is
that the default BSP package (i.e. interfacing IP blocks) costs logic
cells and RAM blocks but is not using any DSP blocks and is incurred
only once, independent of the number of instantiated kernels. 

\subsection{Latency Evaluation}

We experimentally evaluated all the combinations of Table~\ref{tab:Resource-usage-of}
and their execution latencies are given in Table~\ref{tab:latency-combination}.
Only the single configuration file approach of Section~\ref{subsec:Multiple-Devices}
is employed in this section. The recorded values are the execution
latencies of processing one input array ($2^{21}$ complex SPF points)
while applying 42 FIR filters (half of the FOP). Both serial processing
and processing using the multiple buffering technique were evaluated.
For multiple devices, only the multiple buffering-based processing
approach is tested. The major positive observation from the results
in Table~\ref{tab:latency-combination} is that by applying the multiple
buffering technique, the same kernel combination can achieve up to
2x speedup over non-multiple buffering based processing. 

Except for the OLA-256+Na\text{\"i}ve-\textsc{MultipleHP} combination,
all combinations that contain the OLA-TD method apply the OLA-128 kernel.
The value 128 is the largest number of power of two that can be implemented
within the combination. For the combination that contains the \textsc{MultipleHP-R}
method, the FOP preparation module is evaluated by executing on both
host processor or FPGA device(s). 

For combination AOLS+Na\text{\"i}ve-\textsc{MultipleHP} and combination
AOLS+\textsc{MultipleHP}-N, the $t_{HM}$ is larger than $\frac{1}{2}t_{FDAS}$,
so the single configuration file with single input array approach
(in Section~\ref{subsec:Multiple-Devices}) is applied to split the
harmonic-summing task evenly for multiple devices, $\times3$ in this
research, see the last two rows of Table~\ref{tab:latency-combination}.
For the configuration parameters of harmonic-summing kernels, the
applied values are the largest that can be successfully compiled by
AOC for the $A10$ FPGA. For the remaining combinations on three devices,
they all process multiple input arrays in parallel\@.

\begin{wstable}[h]
\begin{centering}
\caption{\label{tab:latency-combination}Execution latency of combined kernels
on $A10$ cards}
\par\end{centering}
{\small{}}%
\centering{}
\begin{tabular}{@{}cccccccc@{}}
\toprule 
\multicolumn{2}{c}{{\small{}FT Convolution}} & {\small{}FOP preparation} & {\small{}Harmonic-summing} & {\small{}No multiple } & {\small{}Multiple } & {\small{}Pipeline} & {\small{}Multiple devices }\tabularnewline
\multicolumn{2}{c}{{\small{}module}} & {\small{}module} & {\small{}module} & {\small{}buffering } & {\small{}buffering} & {\small{}period} & {\small{}$\times3$}\tabularnewline
\multicolumn{2}{c}{} &  &  & {\small{}($ms$)} & type & {\small{}($ms$) } & {\small{}$(ms)$}\tabularnewline
\colrule 
\multirow{7}{*}{\rotatebox{90}{TDFIR}} & {\small{}OLA-128} & {\small{}\textendash{}} & {\small{}Na}\text{\"i}{\small{}ve-}\textsc{\small{}MultipleHP} & {\small{}2,121} & Double & {\small{}1,698 } & {\small{}568}\tabularnewline
 & {\small{}OLA-256} & {\small{}\textendash{}} & {\small{}Na}\text{\"i}{\small{}ve-}\textsc{\small{}MultipleHP} & {\small{}1,278} & Double & {\small{}854} & {\small{}286}\tabularnewline
 & {\small{}OLA-128} & {\small{}transpose} & \textsc{\small{}MultipleHP-N-(1)} & {\small{}2,916} & Double & {\small{}2,219} & {\small{}742}\tabularnewline
 & {\small{}OLA-128} & {\small{}transpose+reorder} & \textsc{\small{}MultipleHP-R-$(16,4)$} & {\small{}3,917} & Double & {\small{}1,935} & {\small{}647}\tabularnewline
 & {\small{}OLA-128} & {\small{}transpose+reorder} & \multirow{2}{*}{\textsc{\small{}MultipleHP-R-$(16,4)$}} & \multirow{2}{*}{{\small{}2,727}} & \multirow{2}{*}{Double} & \multirow{2}{*}{{\small{}2,052}} & \multirow{2}{*}{{\small{}686}}\tabularnewline
 &  & {\small{}(in host)} &  &  &  &  & \tabularnewline
 & {\small{}OLA-128} & {\small{}\textendash{}} & \textsc{\small{}SingleHP-$(S,R,8)$} & {\small{}2,662} & Double & {\small{}1,966} & {\small{}657}\tabularnewline
\multirow{9}{*}{\rotatebox{90}{FDFIR}} & {\small{}AOLS-2048 } & {\small{}discard+transpose} & {\small{}Na}\text{\"i}{\small{}ve-}\textsc{\small{}MultipleHP} & {\small{}856} & Double & {\small{}570} & {\small{}190}\tabularnewline
 & {\small{}AOLS-2048 } & {\small{}discard+transpose} & \textsc{\small{}MultipleHP-N-$(1)$} & {\small{}976} & Double & {\small{}661} & {\small{}224}\tabularnewline
 & {\small{}AOLS-2048 } & {\small{}discard+transpose} & \multirow{2}{*}{\textsc{\small{}MultipleHP-R-$(16,4)$}} & \multirow{2}{*}{{\small{}8,780}} & \multirow{2}{*}{Double} & \multirow{2}{*}{{\small{}6,630}} & \multirow{2}{*}{{\small{}2,219}}\tabularnewline
 &  & {\small{}+reorder} &  &  &  &  & \tabularnewline
 & {\small{}AOLS-2048 } & {\small{}discard+transpose} & \multirow{2}{*}{\textsc{\small{}MultipleHP-R-$(16,4)$}} & \multirow{2}{*}{{\small{}972}} & \multirow{2}{*}{Double} & \multirow{2}{*}{{\small{}633}} & \multirow{2}{*}{{\small{}\textendash{}}}\tabularnewline
 &  & {\small{}+reorder (in host)} &  &  &  &  & \tabularnewline
 & {\small{}AOLS-2048 } & {\small{}discard} & \textsc{\small{}SingleHP-$(S,R,8)$} & {\small{}786} & Double & {\small{}682} & {\small{}237}\tabularnewline
 & {\small{}AOLS-2048 } & {\small{}discard+transpose} & \textsc{\small{}$\frac{1}{3}$ }{\small{}Na}\text{\"i}{\small{}ve-}\textsc{\small{}MultipleHP} & {\small{}$523\times3$} & Triple & {\small{}$307\times3$} & {\small{}$307$}\tabularnewline
 & {\small{}AOLS-2048 } & {\small{}discard+transpose} & \textsc{\small{}$\frac{1}{3}$ MultipleHP-N-$(1)$} & {\small{}$587\times3$} & Triple & {\small{}$334\times3$} & {\small{}$334$}
\tabularnewline
\botrule 
\end{tabular}{\small \par}
\end{wstable}

Except for the FDFIR+\textsc{MultipleHP-R} combination, the FDFIR-based
combinations perform better than these TDFIR-based combinations .
For combinations that contain the OLA-128 kernel, the execution latencies
of the FT convolution part are all around $2s$, which makes them
noncompetitive with FDFIR-based combinations. 

Regarding the FDFIR+\textsc{MultipleHP-R} combination, even though
the \textsc{MultipleHP-R} method is the fastest among the proposed
harmonic-summing methods, the FOP preparation part is inefficient
and the FPGA-based implementation is slower than using the host processor.
It takes $0.6s$ on the host processor and over $8s$ on an $A10$
device, which is over 12x times slower. For the FPGA-based implementation,
the reorder part in the FOP preparation kernel has to leave enough
resources for the main operations and it cannot be parallelised with
a large parallelisation factor. This makes the execution latency of
the FOP preparation part grow up to $8.4s$. If the FOP preparation
task is moved to the host processor, it can process one input array
at a time using all threads, and the pipeline computing on multiple
devices becomes impossible. By considering the FOP reordering, the
advantage of the \textsc{MultipleHP-R} method in low execution latency
disappears.

\subsubsection{Xeon+FPGA (HARP) and GPUs}

The Intel $HARP$ (Xeon+FPGA) platform, as introduced in Section~\ref{sec:FDAS-Module},
supports SVM-based data transfer and it is especially interesting
to evaluate the combinations that require reordering on it. We evaluated
the FDFIR+\textsc{MultipleHP-R} combination (AOLS-2048+\textsc{MultipleHP-R}-$(16,4)$),
and the results are given in Table~\ref{tab:Intel-Xeon}. The same
kernel can achieve higher kernel frequency on $HARP$ than on the
$I7+A10$ system. While the execution latency of each FIR filter of
the FT convolution module on $HARP$ is shorter than on $I7+A10$,
the kernel launching overhead on $HARP$ is higher, which makes the
total execution latency of 42 FIR filters longer than that on $I7+A10$.
The main reason is that the host processor part of the $HARP$ performs
worse than $I7$. Regarding the FOP preparation module, which is processed
on the host processor, the SVM transfer on $HARP$ is about 1.7x times
faster than the general transfer on $I7+A10$. However, the performance
of the host Xeon processor is over 1.6x times worse than that of the
independent $I7$. This weakens the advantage of SVM-based implementation
over the general transfer based implementation. It can be seen that
the overall $t_{FDAS}$ on the $HARP$ platform is 6\% lower than
that on the $I7+A10$. 

\begin{table}
\caption{\label{tab:Intel-Xeon}Comparison of AOLS-2048+\textsc{MultipleHP-R}-$(16,4)$
on $HARP$ and $I7+A10$ device}
\centering{}%
\begin{tabular}{@{}ccccccc@{}}
\toprule 
\multirow{2}{*}{{\small{}Platform}} & {\small{}SVM} & {\small{}Kernel frequency } & {\small{}$t_{FT}$} & {\small{}$t_{FOP}$} & {\small{}$t_{HM}$} & {\small{}$t_{FDAS}$}\tabularnewline
 & {\small{}transfer} & {\small{}$(MHz)$} & {\small{}$(ms)$} & {\small{}$(ms)$} & {\small{}$(ms)$} & {\small{}$(ms)$}\tabularnewline
\colrule 
{\small{}$HARP$} & {\small{}Yes} & {\small{}225.0} & {\small{}347} & {\small{}560} & {\small{}122} & {\small{}1,029}\tabularnewline
{\small{}$I7+A10$} & {\small{}No} & {\small{}185.2} & {\small{}190} & {\small{}633} & {\small{}149} & {\small{}972}\tabularnewline
\botrule 
\end{tabular}
\end{table}

Regarding the $R7$ GPU, since the single work-item kernels such as
candidate detection and FOP preparation are efficient for GPUs, we
only compare the combinations that consists of NDRange kernels to
not distort the result in favour of the FPGAs. The details of the
execution latencies of the GPU-based combinations are given in Table~\ref{tab:vsgpu}.
The parallelisation factors are for the FPGA-based kernels and some
of them do not work for GPU-based kernels. Though $R7$ supports running
multiple kernels currently, the large number of work-groups of FT
convolution kernels and harmonic-summing kernels make it fail to
execute multiple kernels concurrently. It can be seen that the pipeline
period of a single $A10$ is over 1.35x times slower than that
of $R7$, however, three $A10$ cards provider better performance
than $R7$. Also, remember that the $R7$ implementation does less
work as candidate detection is not included.

\begin{table}
\caption{\label{tab:vsgpu}Comparison of NDRange kernels based combinations
between $A10$ and $R7$}
\centering{}{\small{}}%
\begin{tabular}{@{}cccccccc@{}}
\toprule 
{\small{}FT Convolution} & {\small{}FOP preparation} & {\small{}Harmonic-summing} & {\small{}$t_{FT}$} & {\small{}$t_{HM}$} & {\small{}$t_{FDAS}$} & {\small{}Speedup of $A10\times3$}\tabularnewline
{\small{}module} & {\small{}module} & {\small{}module} & {\small{}($ms$)} & {\small{}($ms$) } & {\small{}$(ms)$} & {\small{}over $R7\times1$}\tabularnewline
\colrule 
{\small{}Na\text{\"i}ve-TDFIR} & {\small{}\textendash{}} & {\small{}Na\text{\"i}ve-}\textsc{\small{}MultipleHP} & {\small{}909} & {\small{}36} & {\small{}945} & {\small{}3.33}\tabularnewline
{\small{}Na\text{\"i}ve-TDFIR} & {\small{}\textendash{}} & \textsc{\small{}SingleHP} & {\small{}909} & {\small{}20} & {\small{}929} & {\small{}1.41}\tabularnewline
{\small{}AOLS-2048} & {\small{}(in the host)} & \textsc{\small{}MultipleHP-R (}{\small{}no CD}\textsc{\small{})} & {\small{}67} & {\small{}20} & {\small{}720} & {\small{}1.05}\tabularnewline
\botrule 
\end{tabular}{\small \par}
\end{table}

\subsubsection{Less Filter Coefficients}

Among the TDFIR-based combinations, OLA-256+Na\text{\"i}ve-\textsc{MultipleHP
}is the fastest that is comparable with FDFIR-based combinations.
If the average FIR length $N_{tap}$ can be reduced, the performance
of the TDFIR-based combinations become comparable with those of the
FDFIR-based combinations. The execution performance of combinations
with reduced $N_{tap}$ is given in Table~\ref{tab:Reduce_N_tap}.
Since the experiments so far showed that $t_{FT}$ is dominating in
the FDAS module for the TDFIR-based combinations, the decrease of
$N_{tap}$ directly leads to a reduced execution latency. Still, the
$t_{FT}$ of a TDFIR-based combination, after reducing $N_{tap}$,
might be still be longer than that of FDFIR-based combinations. However,
the sum of $t_{FT}+t_{discard}+t_{transpose}$ for the FDFIR-based
combination can be larger than the $t_{FT}$ of the TDFIR-based combination.

\begin{table}
\caption{\label{tab:Reduce_N_tap}Execution latencies of combinations with
reduced $N_{tap}$ using multiple buffering on three $A10$ devices
$(ms)$}
\centering{}%
\begin{tabular}{@{}cccc@{}}
\toprule 
\multirow{2}{*}{{\small{}$N_{tap}$}} & {\small{}OLA-128+} & {\small{}OLA-128+} & {\small{}OLA-128+}\tabularnewline
 & {\small{}Na\text{\"i}ve-}\textsc{\small{}MultipleHP} & \textsc{\small{}MultipleHP-N-(1)} & \textsc{\small{}SingleHP-$(S,R,8)$}\tabularnewline
\colrule 
{\small{}421} & {\small{}568} & {\small{}742} & {\small{}657}\tabularnewline
{\small{}256} & {\small{}283} & {\small{}370} & {\small{}328}\tabularnewline
{\small{}128} & {\small{}142} & {\small{}226} & {\small{}235}\tabularnewline
\botrule 
\end{tabular}
\end{table}

\subsection{Energy Dissipation and Power Consumption}

In this section, we measure the power consumption and energy dissipation
of the kernels on the $A10$ cards. Since we have no physical access
to the Intel Xeon+FPGA platform, we were not able to measure those
metrics on that platform. Regarding the $A10$ device, we measure
the overall system power consumption in idle status $P_{idle}$, including
FPGA device(s), and the running power $P_{running}$ when the system
is executing kernels on FPGA device(s). The real power consumption
can be achieved by calculating the difference of $P_{running}$ and
$P_{idle}$. When an $A10$ card is installed in the host, it costs
about $20W$ without launching any kernels. In this case, if the kernel
is launched on a single FPGA, only one FPGA device is installed. For
kernels running on three devices, three FPGA acceleration cards are
installed. To make sure the measured $P_{running}$ is stable, each
combination is launched hundreds of times using a loop, which takes
longer than one minute. The power consumption is measured using a
plug-in power meter (Ego smart socket ESS-AU). When a device is configured
with a new bitstream file, the $P_{idle}$ might be changed slightly.
To remove this interference factor, the power consumption of each
combination is measured by 1)~first shutting down the host for a minute
to cool down the host and device(s), 2)~boot the system, and 3) execute
the kernel directly. 

The power consumption and energy dissipation of different types of
combinations on a single $A10$ device are given in Table~\ref{tab:Power-Consumption-Single}.
The overall energy dissipation ($P_{running}\times t_{MB}$) and absolute
energy ($(P_{running}-P_{idle})\times t_{MB}$) dissipation are calculated
based on $P_{idle}$, $P_{running}$, and kernel execution latencies
for one input array, where $t_{MB}$is the latency using the multiple
buffering technique in Table~\ref{tab:latency-combination} (using
42 FIR filters). Based on the number of installed devices the power
consumption in idle status are $P_{idle-FPGA\times1}=49W$ and $P_{idle-FPGA\times3}=89W$.

\begin{table}
\caption{\label{tab:Power-Consumption-Single}Power consumption and energy
dissipation of a single $A10$ device in executing the combined FDAS
module}
\begin{centering}
\begin{tabular}{@{}ccccc@{}}
\toprule 
{\small{}FDAS module} & {\small{}Pipeline} & {\small{}$P_{running}$} & {\small{}Overall} & {\small{}Absolute }\tabularnewline
{\small{}Combinations} & {\small{}computing} &  & {\small{}energy} & {\small{}energy}\tabularnewline
 &  & {\small{}$(W)$} & {\small{}($J$)} & {\small{}($J$)}\tabularnewline
\colrule 
{\small{}FDFIR+}\textsc{\small{}MultipleHP-R} & \multirow{2}{*}{{\small{}No}} & \multirow{2}{*}{{\small{}66}} & \multirow{2}{*}{{\small{}180.2}} & \multirow{2}{*}{{\small{}46.4}}\tabularnewline
{\small{}(FOP preparation in host)} &  &  &  & \tabularnewline
{\small{}FDFIR+}\textsc{\small{}MultipleHP-R} & {\small{}No} & {\small{}59} & {\small{}519.2} & {\small{}88}\tabularnewline
\multirow{2}{*}{{\small{}FDFIR+Na\text{\"i}ve-}\textsc{\small{}MultipleHP}} & {\small{}No} & {\small{}59} & {\small{}50.6} & {\small{}8.6}\tabularnewline
 & {\small{}Yes} & {\small{}60} & {\small{}30.6} & {\small{}6.3}\tabularnewline
\multirow{2}{*}{{\small{}FDFIR+}\textsc{\small{}MultipleHP-N}} & {\small{}No} & {\small{}60} & {\small{}58.6} & {\small{}10.7}\tabularnewline
 & {\small{}Yes} & {\small{}63} & {\small{}35.9} & {\small{}8}\tabularnewline
\multirow{2}{*}{{\small{}FDFIR+}\textsc{\small{}SingleHP}} & {\small{}No} & {\small{}62} & {\small{}48.7} & {\small{}10.2}\tabularnewline
 & {\small{}Yes} & {\small{}64} & {\small{}43.5} & {\small{}10.2}\tabularnewline
\multirow{2}{*}{{\small{}TDFIR+}\textsc{\small{}SingleHP}} & {\small{}No} & {\small{}60} & {\small{}159.7} & {\small{}29.3}\tabularnewline
 & {\small{}Yes} & {\small{}69} & {\small{}135.7} & {\small{}39.3}\tabularnewline
\multirow{2}{*}{{\small{}TDFIR+Na\text{\"i}ve-}\textsc{\small{}MultipleHP}} & {\small{}No} & {\small{}57} & {\small{}120.8} & {\small{}17.0}\tabularnewline
 & {\small{}Yes} & {\small{}59} & {\small{}100.2} & {\small{}17.0}\tabularnewline
\multirow{2}{*}{{\small{}TDFIR+}\textsc{\small{}MultipleHP-N}} & {\small{}No} & {\small{}57} & {\small{}166.2} & {\small{}23.3}\tabularnewline
 & {\small{}Yes} & {\small{}59} & {\small{}130.9} & {\small{}22.2}\tabularnewline
\botrule 
\end{tabular}
\par\end{centering}
\end{table}

The first observation is that the power consumption $P_{running}$
does only vary between $57W$ and $69W$, whereas the energy dissipation
varies significantly more, which is of course due to the large difference
in execution latencies (see Table~\ref{tab:latency-combination}).
For the same combination, the overall energy dissipation by applying
pipeline computing is lower than that without pipeline computing.
Regarding the absolute energy dissipation, applying pipeline computing
costs less energy for most of the combinations\textsc{. }For the TDFIR+\textsc{SingleHP
}and TDFIR+Na\text{\"i}ve-\textsc{MultipleHP} combination\textsc{s,
}the absolute energy dissipation of the pipeline computing based implementations
are about the same as those of without pipeline computing. The reason
is that the longest part of these combinations, which is $max(t_{FT},\,t_{FOP},\,t_{HM})$,
takes a high proportion in the overall execution latency, hence the
pipeline is not balanced enough to provide more benefit. FDFIR+\textsc{SingleHP}
is the only combination that consumes more energy using pipeline computing.
The main reason is that the ratio of $t_{FT}/t_{FDAS}$ is over 75\%,
making the execution latency for a single input array close to the pipeline
period, in other words, the pipelining in not efficient. In addition,
the power consumption of pipeline computing is larger than that 
without pipeline computing, likely due to the need of additional buffers
and the implicit communications and the fact that more processing
is happening at the same time.

When three $A10$ are installed to accelerate the FDAS module, the
$P_{running}$ is about 2x times higher than those using the single
$A10$ device, which is given in Table~\ref{tab:Power-consumptionx3}.
The power consumption of the FDFIR+\textsc{SingleHP} combination is
the highest among these implementations, however, the power consumption
for three $A10$ cards is only $104W$ ($133W-29W$), where $P_{idle-noFPGA}$
is $29W$. It can be found that it is smaller than that of a single
mid-range GPU device, not to mention high-end GPU platforms, which
can cost up to $300W$ per device. For the TDFIR+\textsc{SingleHP}\textsc{\small{}
}combination on GPU (in Figure~\ref{tab:vsgpu}), the power consumption
for one $R7$ card is $97W$, which is larger than the value in Table
\ref{tab:Power-consumptionx3}, which is $88W$ ($117W-29W$).

\begin{table}
\caption{\label{tab:Power-consumptionx3}Power consumption and energy dissipation
of three $A10$ devices in executing the FDAS module combinations
using pipeline computing; energy ratios of using $3\times A10$ over
$1\times A10$ are given in ($\times*$) }
\begin{centering}
\begin{tabular}{@{}cccc@{}}
\toprule
{\small{}FDAS module} & {\small{}$P_{running}$} & {\small{}Overall energy} & {\small{}Absolute energy}\tabularnewline
{\small{}Combinations} & {\small{}$(W)$} & {\small{}($J$)} & {\small{}($J$)}\tabularnewline
\colrule 
{\small{}FDFIR+}\textsc{\small{}SingleHP} & {\small{}133} & {\small{}30.3 ($\times1.4$)} & {\small{}10.4 ($\times1$)}\tabularnewline
{\small{}FDFIR+Na\text{\"i}ve-}\textsc{\small{}MultipleHP} & {\small{}128} & {\small{}24.3 ($\times1.26$)} & {\small{}7.4 ($\times0.85$)}\tabularnewline
{\small{}FDFIR+Na\text{\"i}ve-}\textsc{\small{}MultipleHP} & \multirow{2}{*}{{\small{}126}} & \multirow{2}{*}{{\small{}38.7 }} & \multirow{2}{*}{{\small{}11.4}}\tabularnewline
{\small{}(1/3 of FOP)} &  &  & \tabularnewline
{\small{}FDFIR+}\textsc{\small{}MultipleHP-N} & {\small{}123} & {\small{}27.6 ($\times1.3$)} & {\small{}8.7 ($\times0.92$)}\tabularnewline
{\small{}TDFIR+}\textsc{\small{}SingleHP} & {\small{}117} & {\small{}76.8 ($\times1.77$)} & {\small{}18.4 ($\times2.13$)}\tabularnewline
{\small{}TDFIR+Na\text{\"i}ve-}\textsc{\small{}MultipleHP} & {\small{}108} & {\small{}61.3 ($\times1.63$)} & {\small{}10.8 ($\times1.57$)}\tabularnewline
{\small{}TDFIR+}\textsc{\small{}MultipleHP-N} & {\small{}111} & {\small{}82.1 ($\times1.6$)} & {\small{}16.3 ($\times1.36$)}\tabularnewline
\botrule 
\end{tabular}
\par\end{centering}
\end{table}

By installing three devices, the overall energy dissipation of processing
one input array drops when compared with single device-based processing.
However, the absolute energy dissipations for FDFIR-based combinations
are all increased to some degree (ratio is given in brackets). For
the TDFIR-based combinations, the absolute energy dissipations are
all decreased. For the FDFIR+Na\text{\"i}ve-\textsc{MultipleHP} combination,
the implementation that processes one input array on three devices
(each device processes 1/3 of half FOP) costs more energy than the
implementation that processes three input arrays on three devices.
Although the processing one input array on three devices needs
less power, the same FT convolution and FOP preparation tasks are
redundantly executed three times to avoid communication. Among these
combinations, the absolute costs of FDFIR+Na\text{\"i}ve-\textsc{MultipleHP}
on the single device and three devices are both the smallest, so is
the execution latency in Table~\ref{tab:latency-combination}. 

Regarding the reduction of the average tap number $N_{tap}$, when
$N_{tap}$ is reduced from 421 to 128, the power consumption and energy
dissipation of TDFIR-based combinations are decreased, and the overall
energy consumption is up to 3.9x times less than that of the original
implementation, which is shown in Table~\ref{tab:reduce_tap}.
\begin{table}
\caption{\label{tab:reduce_tap}Power consumption and energy dissipation of
three $A10$ devices in executing the FDAS module combinations using
pipeline computing with reduced $N_{tap}$ ($N_{tap}=128$); energy
ratios of $N_{tap}=421$ over $N_{tap}=128$ are given in ($\times*$)}
\centering{}%
\begin{tabular}{@{}cccc@{}}
\toprule
{\small{}FDAS module} & {\small{}$P_{running}$} & {\small{}Overall energy} & {\small{}Absolute energy}\tabularnewline
{\small{}Combinations} & {\small{}$(W)$} & {\small{}($J$)} & {\small{}($J$)}\tabularnewline
\colrule 
\multirow{2}{*}{{\small{}TDFIR+}\textsc{\small{}SingleHP}} & \multirow{2}{*}{{\small{}119}} & {\small{}27.965} & {\small{}7.11}\tabularnewline
 &  & {\small{}($\times2.7$)} & {\small{}($\times2.6$)}\tabularnewline
\multirow{2}{*}{{\small{}TDFIR+Na\text{\"i}ve-}\textsc{\small{}MultipleHP}} & \multirow{2}{*}{{\small{}114}} & {\small{}25.76} & {\small{}5.65}\tabularnewline
 &  & {\small{}($\times3.2$)} & {\small{}($\times2.9$)}\tabularnewline
\multirow{2}{*}{{\small{}TDFIR+}\textsc{\small{}MultipleHP-N}} & \multirow{2}{*}{{\small{}110}} & {\small{}15.62} & {\small{}2.982}\tabularnewline
 &  & {\small{}($\times3.9$)} & {\small{}($\times3.6$)}\tabularnewline
\botrule 
\end{tabular}
\end{table}

\section{\label{sec:Conclusions}Conclusions}

In paper we have investigated the combination of two well-optimised
pulsar search modules: the FT convolution module and the harmonic-summing
module. We explored the design space of the FDAS module combinations
with different conditions and parallelisation factors using OpenCL.
An FOP preparation module that transforms the FOP based on the demand
of the two neighbouring modules was added to connect them. We also
investigated multiple buffering strategies and assigning the tasks
to multiple devices. As expected, after combining the well-optimised
kernels, the frequency of the combined kernel was slower than any
of its element kernels. The evaluation showed that the method with
the best independent individual performance might not provide good
performance when combined with other modules. Applying the multiple
buffering technique, the combination kernel gains up to 2x times processing
speedup. Among the evaluated combinations, the FDFIR+Na\text{\"i}ve-\textsc{MultipleHP}
performed best and it needed less power and cost less energy than
any other investigated combination. Most of the TDFIR-based combinations
perform worse than the FDFIR-based combinations. When the average
length of the FIR filters can be reduced, the TDFIR-based combinations
showed a high potential in achieving higher performance while costing
less energy. 

\section*{Acknowledgment}

The authors acknowledge discussions with the TDT, a collaboration
between Manchester and Oxford Universities, and MPIfR Bonn and the
work benefitted from their collaboration. We sincerely thank Intel for the donation
of development tools and hardware access, especially access of the HARP systems. 
We gratefully acknowledge that this research was financially supported by the 
SKA funding of the New Zealand government through the Ministry of Business, 
Innovation and Employment (MBIE).



\end{document}